\documentclass[10pt,journal]{IEEEtran}
\usepackage{diagbox}
\usepackage{tikz}
\usepackage{stfloats} 
\UseRawInputEncoding
\usepackage{cite}
\usepackage{amsmath}
\usepackage{amsfonts}
\usepackage{graphicx}
\usepackage{float}

\usepackage[table]{xcolor}
\usepackage{epsfig}
\usepackage{amsthm}
\usepackage{mathrsfs}
\usepackage{amsfonts}
\usepackage{amssymb}
\usepackage{mathrsfs}
\usepackage{euscript}
\usepackage{multirow}
\usepackage{url}
\usepackage{subcaption}
\usepackage{algorithm}
\usepackage{algorithmic}
\usepackage{bm}
\usepackage{mathtools} % for underbracket

\def\LRT#1#2{\!
\raisebox{.2ex}{$
{{\scriptstyle\;#1}\atop{\displaystyle\gtrless}}
\atop
{\raisebox{-1.25ex}{$\scriptstyle\;#2$}}
$}
\!}

\usepackage{xcolor, soul}
\sethlcolor{green}
\def\BibTeX{{\rm B\kern-.05em{\sc i\kern-.025em b}\kern-.08em
    T\kern-.1667em\lower.7ex\hbox{E}\kern-.125emX}}
\theoremstyle{remark}
\theoremstyle{definition}\newtheorem{theorem}{Theorem}

\newtheorem*{remark*}{Remark}

\let\oldproofname=\proofname % this is for bold proof, it needs \usepackage{amsthm}
\renewcommand{\proofname}{\rm\bf{\oldproofname}}

\belowdisplayskip=\abovedisplayskip
\newcommand\submittedtext{%
  \footnotesize This work has been submitted to the IEEE for possible publication. 
  Copyright may be transferred without notice, after which this version may no longer be accessible.}

\newcommand\submittednotice{%
\begin{tikzpicture}[remember picture,overlay]
  \node[anchor=north,yshift=-10pt] at (current page.north)
    {\fcolorbox{red}{white}{%
      \parbox{\dimexpr0.9\textwidth-2\fboxsep-2\fboxrule}{%
    \raggedright\submittedtext}}};
\end{tikzpicture}%
}

\AddToHook{shipout/firstpage}{\submittednotice}

\begin{document}
\bstctlcite{IEEEexample:BSTcontrol}

\title{Target Detection with Tightly-coupled Wideband Antenna Arrays}
\author{Erfan Khordad, \textit{Member}, \textit{IEEE}, Peter J. Smith, \textit{Fellow}, \textit{IEEE}, Sachitha C. Bandara, \textit{Student Member}, \textit{IEEE} and Rajitha Senanayake, \textit{Member}, \textit{IEEE} \thanks{Erfan Khordad, Sachitha C. Bandara and Rajitha Senanayake are with the Department of Electrical and Electronic Engineering, the University of Melbourne, Melbourne, VIC 3052, Australia (e-mail:   khordade@yahoo.com; hbandara@student.unimelb.edu.au; rajitha.senanayake@unimelb.edu.au). 

Peter J. Smith is with the School of Mathematics and Statistics,
Victoria University of Wellington, Wellington 6140, New Zealand (e-mail: smithpe@ecs.vuw.ac.nz). (\textit{Corresponding author: Erfan Khordad})}
}
\maketitle
%\markboth{IEEE COMMUNICATIONS LETTERS,~Vol.~, No.~, ...~2018}%$^{\dag}$
%{Shell \MakeLowercase{\textit{et al.}}: Bare Demo of IEEEtran.cls for Journals}
%\IEEEoverridecommandlockouts
%\IEEEpubid{\makebox[\columnwidth]{978-1-5386-5541-2/18/\$31.00~\copyright2019 IEEE \hfill} \hspace{\columnsep}\makebox[\columnwidth]{ }}

% \markboth{submitted to ??? on \today}
% \IEEEoverridecommandlockouts

\begin{abstract}
This paper focuses on tightly coupled antenna (TCA) array systems with high mutual coupling, resulting in wideband signals. We analyze the target detection capability of such systems with joint time and frequency approaches. We assume a target whose echo signals are unknown wideband signals and show the opportunity that TCAs create in target detection. We consider temporal and spectral dynamics of the signal as either constant, slowly or rapidly varying in frequency and time, giving nine possible cases in total. For each case we derive a detector and present thorough analysis for the distribution of the detector test statistic, considering frequency-dependent noise correlation due to mutual coupling. For the cases with a signal changing slowly in at least one domain we propose a novel detection approach based on estimation of the signal via local averaging of
the samples. We also derive a detector that results in an upper bound on detection performance for comparison. We compare our analytical results with simulations, showing perfect accuracy of our derived distributions. In addition, we assess the performance of the detectors under various time/frequency variation rates. We show the superior performance of our proposed detection approach in comparison to other detectors including an extended energy detector we derive for correlated noise as well as both frequency and time samples.

% Since a wideband spectrum is preferred in wireless communications, target detection with wideband communications waveforms can be embedded into communications systems for the purpose of integrated sensing and communications
\end{abstract}


\section{Introduction}\label{IntroductionLABEL}

Incorporating a large number of antennas in an array is required in beyond-5G devices owing to the demand for increased spectral efficiency. As a result, antennas are placed very closely in a compact space with element spacing smaller than half-wavelength, bringing about non-negligible mutual coupling among antennas \cite{SWmassiveMIMO}. This trend goes even further with continuous-like antennas \cite{RicianContinuousAntenna,Continuous10570543}. 

Mutual coupling (MC) has been considered a negative factor that worsens the performance of communications as well as radar sensing. Therefore, methods were designed to alleviate its impact, e.g. \cite{Arnold1,Jiongpei450}. However, seminal studies including \cite{Munk2006,MunkBook2003} demonstrated that MC can be beneficial. Such merits of MC have not been fully investigated yet. The MC of tightly-coupled antennas (TCAs) that are densely packed contributes to superdirectivity or wide operational bandwidth, but both cannot be achieved at the same time as there is a trade-off between directivity and operational bandwidth \cite{SuperdirectQfactroBW,SuperdirectiveArrays}. We restrict our study to the impact of the wide operational bandwidth offered by TCAs on detection. The large operational bandwidth is shown in practical experiments \cite{PracticalWidebandObserved,Cavallo6374210} and by analysis using physically-consistent modeling that appropriately accounts for the antenna size and MC \cite{SWmassiveMIMO,SachithaConfVTC,SachithaPaper,Bandara_2025_multiuser}. Physically-consistent models are produced from modeling the components of a communications system based on circuit theory, physical laws, electromagnetic theory and communication channel properties \cite{CircuitTheory12,Amico20}. 

% It was shown in \cite{SWmassiveMIMO,SachithaPaper} that physically-consistent models capture the wide operational bandwidth of tightly-coupled antenna arrays, by appropriately accounting for the antenna size and mutual coupling.

Target detection can substantially
benefit from the large operational bandwidth of TCAs, which we demonstrate in this paper. Another important topical application that can benefit from TCAs is integrated sensing and communications (ISAC). The wide bandwidth of TCAs is desirable for such paradigms as both communications and radar sensing gain from large bandwidths. For ISAC systems, target detection is of particular interest, which has been explored without considering physically-consistent models for the antenna elements. For instance, target detection was investigated in \cite{MultiTargetMultiU045,Shayan698} for a multi-target scenario, in \cite{Derrick1625} for wideband signals and in \cite{clutter890} considering the clutter. In these works, various antenna topologies were investigated however without leveraging the merits of MC. 

% the conventional antenna array modeling that lacks physically-consistent considerations was used in the literature.
% Benefits of TCA are yet to be fully investigated in terms of target detection. 

% Wide operational bandwidth tightly-coupled antenna arrays allows for taking wide apart frequency samples for detection.

% There is a rich literature on the detection problem with a variety of methods under various assumptions and system characteristics. Nevertheless, the benefit of tightly-coupled antennas in target detection, considering all possibilities about temporal and spectral dynamics of the signal, has not been investigated. Detection methods that considered MC, wideband models, dynamics of the signal, or joint time and frequency processing are relevant to this paper, hence some of the most relevant works are reviewed in the following.

Outside the context of ISAC, target detection as a standalone problem is well known under a range of assumptions and system characteristics. Detection with unknown MC among the antenna elements was studied in \cite{RadarUnknownMutual12}. However, MC was considered as a cause of performance degradation and a partially known signal model was used.

% and using a secondary set of samples without the signal component.

Energy detectors or multi-channel energy detectors with detectors in multiple sub-bands \cite{MultipleNarrowB_ED} can be used to detect unknown wideband signals. In energy detection, no assumptions are made about the dynamics of the signal. Also energy detection is vulnerable to uncertainty in the noise and is not optimal when the signal is correlated\cite{ED_notOptimal}. Wideband detection was studied in \cite{TimeReversal320} assuming an extended target constituting several reflecting points. In \cite{ED_notOptimal,Peng11095,SharmaRandomCorrelation}, detection was studied by factoring in the dynamics of the signal as a correlation model. In \cite{ED_notOptimal} for a multiple-antenna receiver, the noise was assumed to be independent and identically distributed (i.i.d). Also in \cite{Peng11095}, a single antenna receiver was considered and asymptotic results were provided using the central limit theorem. Moreover, approaches in \cite{Peng11095,SharmaRandomCorrelation,TimeReversal320} considered only white noise. All the above approaches considered detection in either frequency or time domains, making them attractive only for stationary signals, as for non-stationary signals joint frequency and time approaches are required.

% Detection methods have been designed to perform detection of stationary signals either in the time or frequency domains
% \cite{ED_notOptimal} signal correlation was considered
% In \cite{Peng11095} frequency correlation of the signal was considered
% \cite{MultipleNarrowB_ED} was designed for a single antenna receiver and white noise.

For non-stationary signals, joint frequency and time target detection was studied in \cite{Kalra8635078} using an improved version of the Wigner-Ville transform. In \cite{Javed5709749}, energy detection was considered using the spectrogram and Wigner-Ville transform. In \cite{Sattar892667}, a detection method using a filter bank was developed considering colored noise, and the distribution of the detector was approximated using simulated data for a scalar signal. Yet \cite{Kalra8635078,Javed5709749,Sattar892667} lack analysis on the distribution of their detectors.
% [7] time domain, [8] frequency domain, [9] joint time-freq
% several detectors in multiple sub-bands \cite{MultipleNarrowB_ED}, however for a single antenna receiver, uncorrelated signals and white noise. Energy detection was also used in cognitive radio for wideband spectrum sensing but it was mainly used with a different objective than only detection \cite{SurveyCognitive}.

% Performance degradation of the energy detector for correlated signals was investigated in \cite{ED_degradation}

% None of the approaches so far has considered the positive impact of MC in radar detection.

% joint time-frequency energy detectors based on Short time FT and Winger-Ville need to check the enrgy obtained by all observed samples one by one, making it computationly hard?

Nevertheless, the benefits of TCAs for target detection, considering all possibilities about temporal and spectral dynamics of the signal, are still unexplored. As such, in \cite{ErfanConfDetection2025} we studied the target detection problem using TCAs by categorizing the spectral dynamics of the signal into three regimes. We extend the analysis in \cite{ErfanConfDetection2025} to a more complete signal model by factoring in both spectral and temporal variations of the signal, which results in nine cases as explained below.

In this paper, we propose a novel target detection approach. We also show the opportunity that TCAs offer in target detection due to the resultant wide operational bandwidth. Echo signals reflected by the target are assumed to be unknown wideband signals. Thus, any wideband waveform can be used for detection. Since our approach can be used for sensing the target with any wideband waveform, it could be beneficial for both communications-centric and radar-centric ISAC waveform designs.
% such as in \cite{Iceberg11037613} and \cite{Temiz100RadarCentric} respectively. 
Also, MC introduces frequency-dependent noise correlation into the detection problem, which we take into account. We consider the temporal and spectral dynamics of the signal, classifying the dynamics as either constant, slowly or rapidly varying in frequency and time. This gives nine possible cases in total, given in Table \ref{Table:tableFor9Cases}. Specifically, our novel contributions are summarized as follows:
%We consider non-stationary signals to account for a general echo signal model.
\begin{itemize}
\item We consider an unknown wideband signal model in the presence of frequency-dependent correlated noise and take into the account temporal and spectral dynamics of the signal. This results in the nine possible cases discussed above. For each of the cases, we derive a detector and provide a thorough analysis of the distribution of the detector output under null and alternative hypotheses. We show the perfect agreement of our analytical results and simulations.

\item We propose a novel detection approach for the cases with a signal changing slowly in at least one domain. Our approach is based on estimation of the signal by averaging of the samples using a one-dimensional window either in frequency or time, as well as a two-dimensional window in both domains. 

\item We derive a detector that gives an upper bound on the probability of detection, for comparison. In addition, we examine the performance of the nine detectors under various time/frequency variation rates. We derive an extended
energy detector considering correlated noise as well as both frequency and time samples. We show that our proposed detection approach outperforms other detectors including the extended
energy detector. We also highlight the benefit of TCAs over conventional weakly-coupled antennas.
\end{itemize}

\textit{Notation}:
$\chi^2_A(B)$ denotes a non-central chi-square variable with $A$ degrees of freedom (DoFs) and non-centrality parameter, $B$. $\chi^2_A$ denotes a chi-square variable with $A$ DoFs. Non-central chi-square and chi-square variables in the Theorems are assumed to be independent and denoted respectively by $\chi^2_{A(z),z}(B(z))$ and $\chi^2_{A(z),z}$ for a general index $z=1,...,Z$, with $A(z)$ and $B(z)$ being the DoFs and non-centrality parameter. Bold uppercase $\mathbf A$ and bold lowercase $\mathbf a$ denote a matrix and a column vector respectively. Non-bold typeface $A$ denotes a scalar. The conjugate transpose and transpose are denoted by $\mathbf A^H$ and $\mathbf A^T$. The operator $\text{blockdiag}\{\mathbf D_1,\mathbf D_2,...,\mathbf D_N\}$ indicates a block diagonal matrix with sub-matrices $\mathbf D_1$ to $\mathbf D_N$. Summation $\sum_c$ indicates $\sum_{c=c_1}^{c_N}$, for a general index $c=c_1,c_1+1,...,c_N$. Operators $\operatorname{Re}$ and $\operatorname{Im}$ denote the real and imaginary parts of a complex number respectively. Also, $\mathbf 0$ denotes an all-zero vector and $\mathbf I_{N}$ denotes an $N$ by $N$ identity matrix.% Also, $\|\mathbf A\|_F$ denotes the Frobenius norm, $\|\mathbf a\|_2$ denotes 2-norm and $\mathbb{E}$ denotes expectation. $diag(\mathbf A)$ denotes a column vector of diagonal entries of $\mathbf A$. The vectorization of $\mathbf A$ is denoted by $vec(\mathbf A)$.

\section{System Model and Problem Formulation}\label{Sec:SystemModel}

Our proposed detection approach and analysis in this paper are general in the sense that they can apply to any similar joint time and frequency detection problem with frequency varying correlated noise. However, we aim to unveil the merits of tight coupling in target detection. Thus we present a system model using circuit-theoretic parameters and pose the target detection problem for TCAs.

We assume the use of canonical minimum scattering (CMS) antennas that provide a broad bandwidth as they have only the first radiation mode with a low Q-factor. As mentioned in Section \ref{IntroductionLABEL}, TCA arrays can also achieve superdirectivity, but at the cost of larger Q-factor and reduced bandwidth \cite{SuperdirectQfactroBW,SuperdirectiveArrays}.

% Since we focus our attention on the wideband behavior of TCA arrays, CMS antennas are used to model the antenna array.

Our model is based on Chu's theory, and the antennas are referred to as Chu's CMS antennas \cite{SWmassiveMIMO,CMS_Antennas1965}. Based on Chu's theory, an antenna that is restricted in a sphere with a specific radius is modeled with an equivalent resistor-inductor-capacitor (RLC) circuit. Impedance values of the components of the RLC circuit are dependent on the radius. Using Chu's model allows us to account for the impact of MC on the detection performance.

The mutual impedance between two Chu's CMS antennas is given in \cite[Eq. (5)]{SWmassiveMIMO} which was derived by establishing an equivalence between Hertz dipoles and Chu's CMS antennas based on their power radiation. The expression takes into account the antenna spacing, which we denote by $\delta$, and the orientation of the dipoles. We adopt a colinear configuration for the dipoles which is a commonly-used configuration where the dipoles are put along the axis connecting the dipoles.

We consider a transmitter sending wideband signals which are reflected from a target and the echoes arrive at a sensing receiver equipped with a uniform linear array of $N_R$ Chu's CMS antennas. We denote by $a_R$ the radius of the sphere within which a receiver antenna is enclosed. The antenna spacing for the array at the sensing receiver is assumed to be small so that there is tight coupling among the antennas, which results in a TCA array. 

Wideband signals are used to leverage the wideband TCA array for target detection. Echoes from the target are assumed to be unknown wideband signals. Thus, any wideband waveform can be used for detection and we do not impose any other assumptions on the waveform.

% parallel configuration where the dipoles are perpendicular to the axis connecting the dipoles

Considering the equivalent circuit for a Chu's CMS antenna, the sensing receiver's antenna array is modeled as a network of RLC circuits, resulting in a circuit-theoretic model\cite{SWmassiveMIMO}. As such, the received signal is a vector of voltages $\mathbf{v}_{k,t}$ where $k=1,..,K$ and $t=1,...,T$ respectively denote $k$th frequency sample and $t$th time sample,
% In Chu's equivalent RLC circuit model the capacitor and the conductor impedance are given as .... where $R$ denotes the resistor value
with $K$ and $T$ denoting the total number of frequency and time samples, respectively, that the receiver collects.
% Also, denote by $T$ and $K$ the total number of time samples and the total number of frequency domain samples respectively. The radar receiver collects $I=TK$ time and frequency samples in total.

% We assume a small number of time samples so that the signal and noise statistics are constant with respect to time and only change with frequency. Thus a few time samples are taken to collect more energy. Hence the received voltage vector varies by time and frequency and the detection problem is posed as
Our objective is to detect the target using joint time and frequency samples at the sensing receiver. Hence, the detection problem is posed as
\begin{equation}\label{Eq:originalDetProblem}
\begin{split}
    \mathcal{H}_0&: \mathbf v_{k,t} = \mathbf n_{k,t}\\    
    \mathcal{H}_1&: \mathbf v_{k,t} = \bm \alpha_{k,t}+\mathbf n_{k,t},
\end{split}
\end{equation}
with $\mathcal{H}_0$ and $\mathcal{H}_1$ being the null and alternative hypotheses respectively, where vectors $\bm \alpha_{k,t}$ and $\mathbf n_{k,t}$ denote the signal and noise components. The signal component, that is assumed to be unknown, models the echo signal from the target that has been impacted by the channel between the target and the sensing receiver. We employ a physically-consistent model for the channel in Section \ref{Sec:NumericalResults} where we present numerical results. The noise component has a complex correlated Gaussian distribution that models the background noise as well as noise due to low-noise amplifiers and the receive circuit, with frequency-dependent complex covariance matrix $\mathbf{R}_k$ given in \cite[Eq. (16b)]{SWmassiveMIMO}. We assume noise is uncorrelated in time. The hypothesis testing problem in \eqref{Eq:originalDetProblem} is a physically-consistent target detection problem because the channel model and matrix $\mathbf R_k$ both capture the impact of $\delta$ and $a_R$ as well as MC.

The signal $\bm\alpha_{k,t}$ in frequency and time domains can be either constant, slowly or rapidly varying, which we denote by numeric representations 1, 2 and 3 respectively. This gives 9 possible cases in total. We denote such cases by a numeric pair $(a,b)$ where $a$ and $b$ denote the numeric representations in frequency and time respectively. Table \ref{Table:tableFor9Cases} summarizes the cases with their corresponding sections in this paper.

% For example, numeric pair (2,3) indicates that the signal is slowly changing in frequency and rapidly varying in time. We use the numeric pairs as subscripts for detectors corresponding to the 9 possible cases. 

\begin{table}[]
\caption{Nine Cases for Target Detection}\label{Table:tableFor9Cases}
\centering
\addtolength{\tabcolsep}{-4pt}
\begin{tabular}{|l|l|l|l|}
\hline
 \diagbox{Time}{Frequency} & Constant  & Slowly Varying & Rapidly Varying \\ \hline
 Constant & (1,1), Sec \ref{Sec:subSecConstantInBothD} & (2,1), Sec \ref{Sec:subsection2_1_slowFconsT} & (3,1), Sec \ref{Sec:DetectorCase{(3,1)}RapidFreqConsTime} \\ \hline
 Slowly Varying & (1,2), Sec \ref{Subsec:PropForCase(1,2)} & (2,2), Sec \ref{Sec:Subsection_BothSlowlyVarying} & (3,2), Sec \ref{Sec:Case(3,2)} \\ \hline
  Rapidly Varying & (1,3), Sec \ref{Sec:ConstantDetector} & (2,3), Sec \ref{Subec:ProposeDetectorCase(2,3)SlowFRapidT} & (3,3), Sec \ref{Sec:BaselineWideBdetector} \\ \hline
\end{tabular}
\end{table}

In the following, we first review how probabilities of false alarm and detection are calculated for a general detector. We then provide a detection framework and explain how to estimate the signal. In the sections that follow, we derive the 9 particular detectors and provide analysis.

\subsection{Preliminaries: Probabilities of False Alarm and Detection}\label{Sec:preliminaries}

% We will see later that the detectors follow chi-squares or a linear combination of chi-squares under the null or alternative hypotheses.

Calculating probabilities of false alarm and detection is reviewed here for a general detector, given that the distributions of the detector test statistic are known under the null and alternative hypotheses. In Section \ref{Sec:NumericalResults} for the numerical results, the method explained here is used to compute probabilities of false alarm and detection for all the detectors.

Denote by $\Lambda$ a general detector and denote by $\gamma_g$ a general test threshold for $\Lambda$. Also, denote the probability density function (PDF) and cumulative distribution function (CDF) of the detector under the null hypothesis respectively by $f_{\Lambda|\mathcal{H}_0}(x)$ and $F_{\Lambda|\mathcal{H}_0}(x)$. Similarly, denote the PDF and CDF of the detector under the alternative hypothesis respectively by $f_{\Lambda|\mathcal{H}_1}(x)$ and $F_{\Lambda|\mathcal{H}_1}(x)$. The probability of false alarm is calculated as $P_{FA}=\int_{\gamma_g}^\infty f_{\Lambda|\mathcal{H}_0}(x) d x$. Thus for a fixed $P_{FA}$, the test threshold $\gamma_g$ is calculated as $\gamma_g=F_{\Lambda|\mathcal{H}_0}^{-1}(1-P_{FA})$. Also, the probability of detection is computed by $P_{D}=\int_{\gamma_g}^\infty f_{\Lambda|\mathcal{H}_1}(x) d x$ and hence for a fixed $P_{FA}$, it is given as $P_{D} = 1-F_{\Lambda|\mathcal{H}_1}(\gamma_g)$.

\subsection{LLR for Joint Frequency and Time Samples}
We derive a general detection framework in this subsection with no prior assumption about the variation of the signal in frequency or time, based on the log-likelihood ratio (LLR). Denote by $\mathbf V$ the array of all received samples $\mathbf{v}_{k,t}$ in time and frequency. We write the likelihood ratio test (LRT) with threshold $\gamma$ for the detection problem in (\ref{Eq:originalDetProblem}) next. Using the PDF of a correlated Gaussian distribution, the LRT is written as
\begin{equation}\label{Eq:LRT_movingAve1}
\Gamma_\text{LRT} (\mathbf{V}) \!=\!\!\frac{\exp\! \left(\sum_{k} \! \sum_{t} \!\! -(\!\mathbf{v}_{k,t} \!-\! \bm{\alpha}_{k,t}\!)^H \mathbf{R}^{-1}_k  (\!\mathbf{v}_{k,t} \! - \! \bm{\alpha}_{k,t}\!)\!\right)}{ \exp \left(\sum_{k}  \sum_{t} \! -\mathbf{v}_{k,t}^H \mathbf{R}^{-1}_k  \mathbf{v}_{k,t}\right)}\!\! \LRT{\mathcal{H}_1}{\mathcal{H}_0} \!\!\gamma.
\end{equation}
% where $1 \leq k \leq K$ and $1 \leq t \leq T$. The log-likelihood ratio of (\ref{Eq:LRT_movingAve1}) then gives the LLR with threshold $\gamma_1$ as
% \begin{equation}\label{Eq:logLikelihoodForProposedDet}
% \begin{split}
% &\Lambda_\text{LLR} (\mathbf{V}) \\&\!\!= \!\!\sum_{k} \!\! \sum_{t} \!\! \left(\!\mathbf{v}^H_{k,t}\mathbf{R}^{\!-1}_k\!\bm{\alpha}_{k,t}\!+\!\bm{\alpha}^H_{k,t}\!\mathbf{R}^{\!-1}_k\mathbf{v}_{k,t}
% \!-\!\bm\alpha^H_{k,t}\mathbf{R}^{\!-1}_k\!\bm\alpha_{k,t} \! \right) \!\!\!\LRT{\mathcal{H}_1}{\mathcal{H}_0} \!\gamma_1\!.
% \end{split}
% \end{equation}

% \subsection{Estimating the Unknown signal}

Since the signal is unknown, \eqref{Eq:LRT_movingAve1} cannot be used directly. Hence, we use the generalized likelihood ratio test (GLRT) method, where the unknown signal is replaced by its maximum likelihood estimate (MLE). The GLRT method is general and we will use it in Section \ref{Sec:SignalWithoutSlowVar} for signals without slow variations in either of the domains, i.e. Cases (1,1), (1,3), (3,1) and (3,3). Denote the MLE estimate of $\bm{\alpha}_{k,t}$ by $\bm{\alpha}_{k,t}^\text{MLE}$. By taking the log of \eqref{Eq:LRT_movingAve1} and replacing $\bm{\alpha}_{k,t}$ with $\bm{\alpha}_{k,t}^\text{MLE}$, the test statistic for these cases is given as an LLR expression, with threshold $\gamma_1$, as
\begin{equation}\label{Eq:logLikelihoodForGLRTwithMLEbased1}
\begin{split}
\Lambda_\text{GLRT} (\mathbf{V}) \!\! &= \!\!\sum_{k} \!\! \sum_{t} \!\! \bigg(\!\mathbf{v}^H_{k,t}\mathbf{R}^{\!-1}_k\!\bm{\alpha}_{k,t}^\text{MLE}\!+\!({\bm{\alpha}_{k,t}^\text{MLE}})^H\!\mathbf{R}^{\!-1}_k\mathbf{v}_{k,t}
-\\&({\bm{\alpha}_{k,t}^\text{MLE}})^H\mathbf{R}^{\!-1}_k\!\bm{\alpha}_{k,t}^\text{MLE}\! \bigg) \!\LRT{\mathcal{H}_1}{\mathcal{H}_0} \!\gamma_1\!.
\end{split}
\end{equation}

For the other cases, where the signal is slowly changing in at least one domain, finding $\bm{\alpha}_{k,t}$ depends on the precise model for the way in which $\bm{\alpha}_{k,t}$ changes in time or frequency. Such precise knowledge is very hard to obtain and restricts the detection to certain scenarios. Ideally, we would have an estimate of $\bm{\alpha}_{k,t}$ which is useful over a wide range of unknown signal types. A non-parametric estimator is the obvious choice and in particular smoothing estimates are simple, general-purpose and effective. Define $\underline{\bm{\alpha}}_{k,t}$ as an estimate of $\bm{\alpha}_{k,t}$ based on local averaging (LA) for which we give expressions in Section \ref{Sec:ProposedForAtleastOneSlow2}. Hence, using the LLR of (\ref{Eq:LRT_movingAve1}), we create a new test statistic as follows
\begin{equation}\label{Eq:OurNewTestStatisticMovingAVE1}
\begin{split}
\Lambda_\text{LA} (&\mathbf{V}) \\&\!\!= \!\!\sum_{k} \!\! \sum_{t} \!\! \bigg(\!\mathbf{v}^H_{k,t}\mathbf{R}^{\!-1}_k\!\underline{\bm{\alpha}}_{k,t}\!+\!\underline{\bm{\alpha}}_{k,t}^H\!\mathbf{R}^{\!-1}_k\mathbf{v}_{k,t}
\!-\!\underline{\bm{\alpha}}_{k,t}^H\mathbf{R}^{\!-1}_k\!\underline{\bm{\alpha}}_{k,t} \! \bigg) \!\!\!\LRT{\mathcal{H}_1}{\mathcal{H}_0} \!\gamma_1\!.
\end{split}
\end{equation}
In the following section, we present our proposed detection approach using \eqref{Eq:OurNewTestStatisticMovingAVE1}.

\section{Proposed Approach for Cases with a Slowly-Varying Signal in at Least One Domain}\label{Sec:ProposedForAtleastOneSlow2}

We propose a novel detection approach based on the test statistic in \eqref{Eq:OurNewTestStatisticMovingAVE1}, for all the cases in which the signal is varying slowly in at least one of the time or frequency domains and present analysis for them. These cases are Case (1,2), Case (2,1), Case (2,2), Case (2,3) and Case (3,2).

We take into consideration the limited knowledge about the slow variation property of the signal in the development of an enhanced detection approach. Our proposed approach is based on LA of the received voltage vectors in the domain where the signal is slowly changing. We adopt a window that slides through time and/or frequency and uniformly averages the samples within it, to estimate the signal using surrounding time and/or frequency samples. We use uniform averaging but the averaging could be non-uniform in general. We show by numerical results in Section \ref{Sec:NumericalResults} that our proposed detectors that factor in signal variation outperform the MLE-based detectors, derived in Section \ref{Sec:SignalWithoutSlowVar}, that lack the ability to use the information about signal variation.

%added 15 Nov 2026: MLE based designed for a specific scenario like for exampe (1,3) freq constant and time rapid, if scenario changes it's not working well, but ours work better when signal rate of variation changes. So ours are not only for one particular scenario.

We show that our proposed detectors under the null and alternative hypotheses are linear combinations of non-central chi-square and chi-square variables that are special cases of a \textit{generalized} chi-square. The cumulative distribution function (CDF) of the generalized chi-square is efficiently generated with software using the Imhof method \cite{AbhranilComputeGeneralized,Abhranil2}.
% A generalized chi-square variable is a sum of weighted chi-square variables, a scaled Gaussian variable and a constant.

Note that \cite{Peng11095} and \cite{DyaynSlidingWin} also used a window sliding through samples in the frequency and time domains respectively. However, the detection in \cite{DyaynSlidingWin} was for a known signal with uncorrelated noise and \cite{Peng11095} studied a special case as mentioned in Section \ref{IntroductionLABEL}.

\subsection{Proposed Detector for Case (1,2): Constant in Frequency and Slowly-varying in Time}\label{Subsec:PropForCase(1,2)}

We make use of the slow variation in time and use a window in the time domain. As the signal is constant in frequency, for the ease of design and analysis we drop the index $k$ from $\bm \alpha_{k,t}$. The window slides only in the time domain and all the frequency bins are averaged uniformly. We estimate $\bm \alpha_t$ with $\bm {\alpha}_t^{(1,2)}$, which is a weighted average of received voltage vectors in the time domain as
\begin{equation}\label{Eq:windowingFormulaCase(1,2)FixFreqSlowTime}
\bm {\alpha}_t^{(1,2)}=\sum_{k} \sum_{{\ell_1}=-{L_1}}^{L_1} \mathbf{v}_{k,t-{\ell_1}} w_{k,t,{\ell_1}},   
\end{equation}
where $w_{k,t,{\ell_1}}$ are the weights of the window and $L_1$ determines the length of the window, i.e., $2L_1+1$. For the averaging, the center of the window with $\ell_1=0$ lies on the $t$-th time sample. We calculate the weights based on uniform averaging of the voltage vectors, but \eqref{Eq:windowingFormulaCase(1,2)FixFreqSlowTime} is general and any weighting approach can be utilized. To avoid edge effects, $w_{k,t,{\ell_1}}$ is set to zero when $t-{\ell_1}<1$ or $t-{\ell_1}>T$. In these edge cases, the averaging is restricted to time samples within the range $1\leq t \leq T$.
% When $t$ and $\ell_1$ satisfy $1\leq t-\ell_1 \leq T$, vector $\mathbf{v}_{k,t-{\ell_1}}$ exists and due to this when $t$ is small or large, some of the weights $w_{k,t,\ell_1}$ cannot have a value as their corresponding vectors $\mathbf{v}_{k,t-{\ell_1}}$ do not exist. This pertains to edge cases where the center of the window lies on starting or ending time samples and some part of the window goes below or beyond time samples, i.e., less than $t=1$ and larger than $t=T$ respectively. In these edge cases, a part of the window overlaps with the time samples and the uniform averaging is done for that part, indicating the signal is estimated by a smaller window or a smaller number of samples. Except for the edge cases, the other condition is when the window fully overlaps with the samples. As a result, there are three conditions for the weights.
Also for $\ell_1<-L_1$ or $\ell_1>L_1$, $w_{k,t,{\ell_1}}$ is set to zero. Hence, we obtain the weights $w_{k,t,\ell_1}$ as
\begin{equation}\label{Eq:Three_casesForWinCase(1,2)FixFreqSlowTime}
w_{k,t,\ell_1}\!\!=\!\!
\begin{cases}
\! 1/\!(K(2L_1+1)) \hspace{1.3cm}   t\geq \!L_1+\!1, t\leq \! T\!-\!L_1\\
\! 1/\!(K(t+L_1)) \hspace{1.5cm} t\leq L_1 \\
\! 1/\!(K(T-t+1+L_1)) \hspace{0.2cm}  t\geq T-L_1+1,
\end{cases}
\end{equation}
where we calculate the weights based on uniform averaging across both domains, with the constraint $\sum_k \sum_{{\ell_1}=-{L_1}}^{L_1} w_{k,t,{\ell_1}} =1$. For the frequency domain, we take the average of all the frequency bins uniformly. Hence, for any $k$ we have $\sum_{{\ell_1}=-{L_1}}^{L_1} w_{k,t,{\ell_1}} =1/K$. Note that calculation of weights for other cases in this section, i.e. Case (2,1), Case (2,2), Case (2,3) and Case (3,2), is also performed piecewise, considering the edge effects.

Our proposed detector for Case (1,2) is written by replacing (\ref{Eq:windowingFormulaCase(1,2)FixFreqSlowTime}) in (\ref{Eq:OurNewTestStatisticMovingAVE1}) as
\begin{equation}\label{Eq:PropDetectorForCase(1,2)FirstEquation}
\begin{split}
&\Lambda_{{(1,2)},\text{p}} (\mathbf{V}) =\\
& \sum_{t} \Bigg[\Bigg(\sum_{k} \mathbf v_{k,t}^H \mathbf{R}^{-1}_k\Bigg) \Bigg(\sum_{k'} \sum_{{\ell_1}=-{L_1}}^{L_1}   \mathbf{v}_{k',t-{\ell_1}} w_{k',t,{\ell_1}}\Bigg) \Bigg]\\
& + \sum_{t} \Bigg[ \Bigg(\sum_{k'} \sum_{{\ell_1}=-{L_1}}^{L_1}   \mathbf{v}_{k',t-{\ell_1}}^H w_{k',t,{\ell_1}}\Bigg) \Bigg(\sum_{k} \mathbf{R}^{-1}_k \mathbf v_{k,t} \Bigg) \Bigg]\\
& - \sum_{t} \Bigg[ \Bigg(\sum_{k'} \sum_{{\ell_1}=-{L_1}}^{L_1}   \mathbf{v}_{k',t-{\ell_1}}^H w_{k',t,{\ell_1}}\Bigg) \Bigg(\sum_{k} \mathbf{R}^{-1}_k \Bigg) \\ &
\Bigg(\sum_{\tilde{k}} \sum_{{\tilde{\ell}_1}=-{L_1}}^{L_1}  
\mathbf{v}_{\tilde{k},t-{\tilde{\ell}_1}} w_{\tilde{k},t,{\tilde{\ell}_1}}\Bigg) \Bigg],
\end{split}
\end{equation}
where subscript $\text{p}$ indicates our proposed detector. After some algebra, we write \eqref{Eq:PropDetectorForCase(1,2)FirstEquation} in a quadratic from as
\begin{equation}\label{Eq:PropDetectorForCase(1,2)2ndEquationQuadratic}
\Lambda_{{(1,2)},\text{p}} (\mathbf{V}) =\sum_{k_1=1}^K\sum_{t_1=1}^T\sum_{k_2=1}^K\sum_{t_2=1}^T \mathbf{v}_{k_1,t_1}^H \mathbf A_{k_1t_1k_2t_2}  \mathbf{v}_{k_2,t_2},
\end{equation}
where we calculate matrix $\mathbf A_{k_1t_1k_2t_2}$ as
\begin{equation}\label{Eq:MatrixAk1t1k2t2ForCAse(1,2)}
\begin{split}
\mathbf A_{k_1t_1k_2t_2} =& \mathbf{R}^{-1}_{k_1} w_{k_2,t_1,{t_1-t_2}}  + w_{k_1,t_2,{t_2-t_1}} \mathbf{R}^{-1}_{k_2} 
\\& - \sum_{t} w_{k_1,t,{t-t_1}} \Bigg(\sum_{k} \mathbf{R}^{-1}_k \Bigg) w_{k_2,t,{t-t_2}}.
\end{split}
\end{equation}
% where since $-L_1\leq \ell_1 \leq L_1$, the first term exists when $-L_1\leq t_1-t_2 \leq L_1$, the second term exists if $-L_1\leq t_2-t_1 \leq L_1$ and the third term exists when $-L_1\leq t-t_1 \leq L_1$ and $-L_1\leq t-t_2 \leq L_1$, otherwise any term that does not satisfy its corresponding condition is set to a zero matrix.

We further simplify our proposed detector in (\ref{Eq:PropDetectorForCase(1,2)2ndEquationQuadratic}) and present it in a matrix form. Define a $KTN_R$ by $KTN_R$ block matrix $\mathbf{\overline{A}}$. Block $(i,j)$ in $\mathbf{\overline{A}}$ is $\mathbf A_{k_1t_1k_2t_2}$ where the indices are
\begin{equation}\label{Eq:IndicesInMatrixA_Bar_k1t1k2t2}
\begin{split}
&k_1=\left\lfloor (i-1)/T \right\rfloor+1,\,\,\, t_1=i-(k_1-1)T,\\
&k_2=\left\lfloor (j-1)/T \right\rfloor+1,\,\,\, t_2=j-(k_2-1)T,
\end{split}
\end{equation}
% \begin{equation}\label{Eq:DefineMatrixAbarForCase(1,2)}
% \mathbf{\overline{A}} = \begin{pmatrix}
% \mathbf A_{1111},\mathbf A_{1112},...,\mathbf A_{111T},...,\mathbf A_{11K1},...,\mathbf A_{11KT}\\
% \mathbf A_{1211},\mathbf A_{1212},...,\mathbf A_{121T},...,\mathbf A_{12K1},...,\mathbf A_{12KT}\\
% \vdots  \\
% \mathbf A_{1T11},\mathbf A_{1T12},...,\mathbf A_{1T1T},...,\mathbf A_{1TK1},...,\mathbf A_{1TKT}\\
% \vdots  \\
% \mathbf A_{K111},\mathbf A_{K112},...,\mathbf A_{K11T},...,\mathbf A_{K1K1},...,\mathbf A_{K1KT}\\
% \vdots  \\
% \mathbf A_{KT11},\mathbf A_{KT12},...,\mathbf 
% A_{KT1T},...,\mathbf A_{KTK1},...,\mathbf A_{KTKT}\\
% \end{pmatrix}
% \end{equation}
with $i=1,2,...,KT$ and $j=1,2,...,KT$. Define a $KTN_R$ by 1 vector $\mathbf {\overline{v}}$ as  
\begin{equation}\label{Eq:DefineMatrixV_Bar_ForCase(1,2)}
\mathbf {\overline{v}} =
\begin{bmatrix}
\mathbf v^T_{1,1},...,\mathbf v^T_{1,T}, \mathbf v^T_{2,1}, ... ,\mathbf v^T_{2,T},..., \mathbf v^T_{K,1}, ... , \mathbf v^T_{K,T}
\end{bmatrix}^T.   
\end{equation}
Therefore, our proposed detector in (\ref{Eq:PropDetectorForCase(1,2)2ndEquationQuadratic}) is written as
\begin{equation}\label{Eq:PropDetectorCase(1,2)MatrixForm}
\Lambda_{{(1,2)},\text{p}} (\mathbf{V}) = \mathbf {\overline{v}}^H \mathbf {\overline{A}}\,\, \mathbf {\overline{v}}.
\end{equation}

Now we provide an analysis of the proposed detector for Case (1,2) in the following theorem.

\begin{theorem}\label{Theorem:TheoremForCase(2,1)ConstantFreqSlowTtime}
The detector $\Lambda_{(1,2),\text{p}}(\mathbf{V})$ in (\ref{Eq:PropDetectorCase(1,2)MatrixForm}) under $\mathcal{H}_1$ and $\mathcal{H}_0$ is respectively a linear combination of non-central chi-squares and a linear combination of chi-squares, as
\begin{equation}\label{Eq:InTheoremForPDforCase(1,2)ConstantFreqSlowTtime}
\Lambda_{(1,2),\text{p}}|\mathcal{H}_1=  \sum_{j'=1}^{KTN_R} \overline{\lambda}_{j'} \chi^2_{2,j'}\bigg( \operatorname{Re}\{\overline{\mu}_{j'}\}^2+\operatorname{Im}\{\overline{\mu}_{j'}\}^2\bigg),
\end{equation}
\begin{equation}\label{Eq:InTheoremForP_FAforCase(1,2)ConstantFreqSlowTtime}
\Lambda_{(1,2),\text{p}}|\mathcal{H}_0= \sum_{j'=1}^{KTN_R} \overline{\lambda}_{j'} \chi^2_{2,j'},
\end{equation}
where $\overline{\lambda}_{j'}$ and $\overline{\mu}_{j'}$ are defined above \eqref{Eq:DefineCOverlineInCase(1,2)} and below \eqref{Eq:DefineMU_Overline_InCase(1,2)} respectively.
\end{theorem}
\begin{proof}
We designed the detector for Case (1,2) when the signal is constant over frequency, but in practice detectors are applied where the assumptions are reasonable but not exact. Hence, we consider a general scenario where the signal is changing in frequency and time. Consequently, to keep our analysis general for Case (1,2) we include the frequency index $k$ and we use $\bm{\alpha}_{k,t}$ in our analysis. Similarly for other detectors we conduct analysis for a general scenario with a signal changing in both domains.

Under $\mathcal{H}_1$, $\mathbf{v}_{k,t}$ in (\ref{Eq:originalDetProblem}) has a Gaussian distribution with mean $\bm \alpha_{k,t}$ and covariance matrix $\mathbf{R}_k$. Hence, we can write $\mathbf v_{k,t} = \bm{\alpha}_{k,t}+\mathbf{R}_k^\frac{1}{2}\tilde{\mathbf{v}}_{k,t}$ where $\tilde{\mathbf{v}}_{k,t} \sim \mathcal{CN}(\mathbf 0,\, \mathbf I)$. Hence, for the alternative hypothesis we substitute $\bm{\alpha}_{k,t}+\mathbf{R}_k^\frac{1}{2}\tilde{\mathbf{v}}_{k,t}$ for $\mathbf{v}_{k,t}$ in (\ref{Eq:PropDetectorCase(1,2)MatrixForm}) to calculate $\Lambda_{(1,2),\text{p}}|\mathcal{H}_1$. Define 
\begin{equation}\label{Eq:DefineVectorb_Bar_ForCase(1,2)}
\mathbf {\overline{b}} =
\begin{bmatrix}
\tilde{\mathbf b}^T_{1,1},...,\tilde{\mathbf b}^T_{1,T}, \tilde{\mathbf b}^T_{2,1}, ... ,\tilde{\mathbf b}^T_{2,T},..., \tilde{\mathbf b}^T_{K,1}, ... , \tilde{\mathbf b}^T_{K,T}
\end{bmatrix}^T\!\!,   
\end{equation}
with $\tilde{\mathbf{b}}_{k,t}=\mathbf{R}_k^{-\frac{1}{2}}\bm{\alpha}_{k,t}+\tilde{\mathbf{v}}_{k,t}$. Also define $\mathbf {\overline{R}}$ as a $KTN_R$ by $KTN_R$ block diagonal matrix, with blocks $\mathbf{R}_k^{\frac{1}{2}}$ for $k=1,...,K$ and each block repeats $T$ times on the diagonal, i.e., 
\begin{equation}\label{Eq:BlockDiagonalMAt_R_BAR_for(1,2)}
\mathbf {\overline{R}}\!=\! \text{blockdiag}\{\mathbf{R}_1^{\frac{1}{2}}\!,...,\!\mathbf{R}_1^{\frac{1}{2}}\!,\mathbf{R}_2^{\frac{1}{2}}\!,...,\mathbf{R}_2^{\frac{1}{2}}\,...,\mathbf{R}_K^{\frac{1}{2}}\!,...,\mathbf{R}_K^{\frac{1}{2}}\!\}.
\end{equation}
Therefore, we give (\ref{Eq:PropDetectorCase(1,2)MatrixForm}) as
\begin{equation}\label{Eq:ProofPropDetectorCase(1,2)First}
\Lambda_{{(1,2)},\text{p}} (\mathbf{V}) = \mathbf {\overline{b}}^H \mathbf {\overline{R}}\,\,\mathbf {\overline{A}}\,\,\mathbf {\overline{R}} \,\, \mathbf {\overline{b}}.
\end{equation}
We replace matrix $\mathbf {\overline{R}}\,\,\mathbf {\overline{A}}\,\,\mathbf {\overline{R}}$ in (\ref{Eq:ProofPropDetectorCase(1,2)First}) with its eigenvalue decomposition as $\bm {\overline{\Phi}} \,\, \bm {\overline{\Delta}} \,\,\bm {\overline{\Phi}}^H$ with $\bm {\overline{\Phi}}$ being a unitary matrix and $\bm {\overline{\Delta}}$ being a diagonal matrix containing the eigenvalues, $\overline{\lambda}_{j'}$. Define
\begin{equation}\label{Eq:DefineCOverlineInCase(1,2)}
 \overline{\mathbf c}\! = \!\overline{\bm{\Phi}}^H \!\!
\begin{bmatrix}
\tilde{\mathbf{v}}^T_{1,1},...,\!\tilde{\mathbf{v}}^T_{1,T},\! \tilde{\mathbf{v}}^T_{2,1}, ... ,\!\tilde{\mathbf{v}}^T_{2,T},..., \!\tilde{\mathbf{v}}^T_{K,1}, ... , \tilde{\mathbf{v}}^T_{K,T}
\end{bmatrix}^T\!\!,   
\end{equation}
and define vector $\overline{\bm{\mu}}$ with $KTN_R$ elements as
\begin{equation}\label{Eq:DefineMU_Overline_InCase(1,2)}
\overline{\bm{\mu}} \! = \! \overline{\bm{\Phi}}^H \!\!\begin{bmatrix}\!
\mathbf p^T_{1,1},...,\!\mathbf p^T_{1,T}, \!\mathbf p^T_{2,1}, ... ,\!\mathbf p^T_{2,T},...,\! \mathbf p^T_{K,1}, ... ,\! \mathbf p^T_{K,T}\!
\end{bmatrix}^T\!\!,   
\end{equation}
with $\mathbf p_{k,t}=\mathbf{R}_k^{-\frac{1}{2}} \bm{\alpha}_{k,t}$. In addition, denote the $j'$-th element of vectors $\overline{\bm{\mu}}$ and $ \overline{\mathbf c}$ respectively by $\overline{\mu}_{j'}$ and $\overline{c}_{j'}$.
Substituting the eigenvalue decomposition into (\ref{Eq:ProofPropDetectorCase(1,2)First}) gives
\begin{equation}\label{Eq:ProofPropDetectorCase(1,2)Second2nd}
\Lambda_{{(1,2)},\text{p}} (\mathbf{V}) =
(\overline{\bm{\mu}}^H + \overline{\mathbf c}^H )\overline{\bm{\Delta}}(\bm{\overline\mu} + \mathbf {\overline{c}} )=\sum_{j'=1}^{KTN_R}\overline{\lambda}_{j'}  |\overline{\mu}_{j'}+\overline{c}_{j'}|^2.
\end{equation}
The summation in (\ref{Eq:ProofPropDetectorCase(1,2)Second2nd}) is a linear combination of $KTN_R$ non-central chi-squares with 2 DoFs and non-centrality parameter $\operatorname{Re}\{\overline{\mu}_{j'}\}^2+\operatorname{Im}\{\overline{\mu}_{j'}\}^2$. This proves (\ref{Eq:InTheoremForPDforCase(1,2)ConstantFreqSlowTtime}). To prove (\ref{Eq:InTheoremForP_FAforCase(1,2)ConstantFreqSlowTtime}), a similar approach can be taken by replacing $\mathbf v_{k,t}$ in (\ref{Eq:PropDetectorCase(1,2)MatrixForm}) with $\mathbf{R}_k^\frac{1}{2}\tilde{\mathbf{v}}_{k,t}$.
\end{proof}

\subsection{Proposed Detector for Case (2,1): Slowly-varying in Frequency and Constant in Time}\label{Sec:subsection2_1_slowFconsT}

We benefit from the slow variation in frequency and use a window in the frequency domain to develop our proposed detector. The signal is constant in time so that $\bm \alpha_{k,t}=\bm \alpha_{k}$. As such, the window moves only in the frequency domain and all samples are averaged uniformly in time. We use an estimate of $\bm \alpha_k$ consisting of the weighted average of the received voltage vectors as
% $\bm \alpha_k^{(2,1)}$ that is a window
\begin{equation}\label{Eq:windowingFormulaCase(2,1)SlowFreqFixedTime}
\bm \alpha_{k}^{(2,1)}=\sum_{t} \sum_{{\ell_2}=-{L_2}}^{L_2} \mathbf{v}_{k-{\ell_2},t} w_{k,t,{\ell_2}}.   
\end{equation}
In \eqref{Eq:windowingFormulaCase(2,1)SlowFreqFixedTime}, $w_{k,t,{\ell_2}}$ are the weights and the window length is  $(2L_2+1)$, with the center of the window at $\ell_2=0$. 

For $\ell_2<-L_2$ or $\ell_2>L_2$, $w_{k,t,{\ell_2}}$ is set to zero. Also since frequency indices are from $1$ to $K$, $w_{k,t,{\ell_2}}$ is set to zero for $k-\ell_2<1$ or $k-\ell_2>K$ to avoid edge effects.
% condition $1\leq k-\ell_2 \leq K$ needs to be satisfied such that vectors $\mathbf{v}_{k-{\ell_2},t}$ exist. 
Thus, there are three conditions for the weights, as in Section \ref{Subsec:PropForCase(1,2)}, and we obtain the weights $w_{k,t,\ell_2}$ as follows
\begin{equation}\label{Eq:Three_casesForWinCase(2,1)SlowFreqConstantTime}
w_{k,t,\ell_1}=
\begin{cases}
1/(T(2L_2+1)) &\!\!\!\!\!\!\!\!\!\!k\geq L_2+1, k\leq K-L_2\\
1/(T(k+L_2)) &\!\!k\leq L_2 \\
1/(T(K-k+1+L_2)) &\!\!k\geq K-L_2+1.
\end{cases}
\end{equation}
We calculated the weights in (\ref{Eq:Three_casesForWinCase(2,1)SlowFreqConstantTime}) by uniform averaging across both time and frequency domains. Hence, $\sum_{t} \sum_{{\ell_2}=-{L_2}}^{L_2} w_{k,t,{\ell_2}} =1$ and owing to uniform averaging in the time domain for any $t$ we also have $\sum_{{\ell_2}=-{L_2}}^{L_2} w_{k,t,{\ell_2}} =1/T$.

Now, using the weighted average in (\ref{Eq:windowingFormulaCase(2,1)SlowFreqFixedTime}) and referring to (\ref{Eq:OurNewTestStatisticMovingAVE1}) our proposed detector can be expressed as
\begin{equation}\label{Eq:PropDetectorForCase(2,1)FirstEquationSlowFFixT}
\begin{split}
&\Lambda_{{(2,1)},\text{p}} (\mathbf{V}) =\\
& \sum_{k} \Bigg(\sum_{t} \mathbf v_{k,t}^H \Bigg) \mathbf{R}^{-1}_k \Bigg(\sum_{t'} \sum_{{\ell_2}=-{L_2}}^{L_2}   \mathbf{v}_{{k-\ell_2},t'} w_{k,t',{\ell_2}}\Bigg)\\
& + \sum_{k} \Bigg(\sum_{t'} \sum_{{\ell_2}=-{L_2}}^{L_2}   \mathbf{v}_{{k-\ell_2},t'}^H w_{k,t',{\ell_2}}\Bigg) \mathbf{R}^{-1}_k \Bigg(\sum_{t}  \mathbf v_{k,t} \Bigg) \\
& - T \sum_{k}  \Bigg(\sum_{t'} \sum_{{\ell_2}=-{L_2}}^{L_2}   \mathbf{v}_{{k-\ell_2},t'}^H w_{k,t',{\ell_2}}\Bigg)  \mathbf{R}^{-1}_k \\ &
\Bigg(\sum_{\tilde{t}} \sum_{{\tilde{\ell}_2}=-{L_2}}^{L_2}  
\mathbf{v}_{k-{\tilde{\ell}_2},\tilde{t}} w_{k,\tilde{t},{\tilde{\ell}_2}}\Bigg).
\end{split}
\end{equation}
Again, some algebra allows us to write \eqref{Eq:PropDetectorForCase(2,1)FirstEquationSlowFFixT} in a quadratic from as follows
\begin{equation}\label{Eq:PropDetectorForCase(2,1)2ndEquationQuadraticSlowFFixT}
\Lambda_{{(2,1)},\text{p}} (\mathbf{V}) \!=\!\sum_{k_1=1}^K\!\sum_{t_1=1}^T\!\sum_{k_2=1}^K\!\sum_{t_2=1}^T \mathbf{v}_{k_1,t_1}^H \mathbf B_{k_1t_1k_2t_2}  \mathbf{v}_{k_2,t_2},
\end{equation}
where $\mathbf B_{k_1t_1k_2t_2}$ is given by
\begin{equation}\label{Eq:MatrixAk1t1k2t2ForCAse(1,2)}
\begin{split}
\mathbf B_{k_1t_1k_2t_2} &= \mathbf{R}^{-1}_{k_1} w_{k_1,t_2,{k_1-k_2}}  + w_{k_2,t_1,{k_2-k_1}} \mathbf{R}^{-1}_{k_2} 
\\& - T\sum_{k} w_{k,t_1,{k-k_1}} \Bigg(\sum_{k} \mathbf{R}^{-1}_k \Bigg) w_{k,t_2,{k-k_2}}.
\end{split}
\end{equation}
% where as $-L_2\leq \ell_2 \leq L_2$, the first term exists when $-L_2\leq k_1-k_2 \leq L_2$, the second term exists if $-L_2\leq k_2-k_1 \leq L_2$ and the third term exists when $-L_2\leq k-k_1 \leq L_2$ and $-L_2\leq k-k_2 \leq L_2$, otherwise any term that does not satisfy its corresponding condition is set to a zero matrix.

We can now write (\ref{Eq:PropDetectorForCase(2,1)2ndEquationQuadraticSlowFFixT}) in a matrix form by defining a $KTN_R$ by $KTN_R$ block matrix $\mathbf{\overline{B}}$. Block $(i,j)$ in $\mathbf{\overline{B}}$ is $\mathbf B_{k_1t_1k_2t_2}$ by indices $k_1$, $t_1$, $k_2$ and $t_2$ which are computed referring to (\ref{Eq:IndicesInMatrixA_Bar_k1t1k2t2}), and $i=1,2,...,KT$ and $j=1,2,...,KT$. Therefore, referring to (\ref{Eq:PropDetectorForCase(2,1)2ndEquationQuadraticSlowFFixT}), we give our proposed detector as
\begin{equation}\label{Eq:PropDetectorCase(2,1)MatrixFormSlowFreqFixedTime}
\Lambda_{{(2,1)},\text{p}} (\mathbf{V}) = \mathbf {\overline{v}}^H \mathbf {\overline{B}}\,\, \mathbf {\overline{v}},
\end{equation}
where $\mathbf {\overline{v}}$ is given in (\ref{Eq:DefineMatrixV_Bar_ForCase(1,2)}). We now formalize our analysis for Case (2,1) in the following theorem.

\begin{theorem}\label{Theorem:TheoremForCase(2,1)SlowFreqFiXedTime}
The detector $\Lambda_{(2,1),\text{p}}(\mathbf{V})$ in (\ref{Eq:PropDetectorCase(2,1)MatrixFormSlowFreqFixedTime}) under $\mathcal{H}_1$ and $\mathcal{H}_0$ is respectively a linear combination of non-central chi-squares and a linear combination of chi-squares, as
\begin{equation}\label{Eq:InTheoremForPDforCase(2,1)SlowFreqFiXedTime}
\Lambda_{(2,1),\text{p}}|\mathcal{H}_1=  \sum_{j'=1}^{KTN_R} \widehat{\lambda}_{j'} \chi^2_{2,j'}\bigg( \operatorname{Re}\{\widehat{\mu}_{j'}\}^2+\operatorname{Im}\{\widehat{\mu}_{j'}\}^2\bigg),
\end{equation}
\begin{equation}\label{Eq:InTheoremForP_FAfor(2,1)SlowFreqFiXedTime}
\Lambda_{(2,1),\text{p}}|\mathcal{H}_0= \sum_{j'=1}^{KTN_R} \widehat{\lambda}_{j'} \chi^2_{2,j'},
\end{equation}
where $\widehat{\lambda}_{j'}$ and $\widehat{\mu}_{j'}$ are defined below \eqref{Eq:ProofPropDetector(2,1)SlowFreqFiXedTime_First}.
\end{theorem}
\begin{proof}
Our proposed detector for Case (2,1) can be used in a general scenario where the signal is time-varying. Thus, for the purpose of generality we consider a time-varying signal. Thus, we use $\bm{\alpha}_{k,t}$ in the analysis.

To calculate $\Lambda_{(2,1),\text{p}}|\mathcal{H}_1$ we substitute $\bm{\alpha}_{k,t}+\mathbf{R}_k^\frac{1}{2}\tilde{\mathbf{v}}_{k,t}$ for $\mathbf{v}_{k,t}$ in (\ref{Eq:PropDetectorCase(2,1)MatrixFormSlowFreqFixedTime}). Therefore, as in Section \ref{Subsec:PropForCase(1,2)}, (\ref{Eq:PropDetectorCase(2,1)MatrixFormSlowFreqFixedTime}) can be expressed as
\begin{equation}\label{Eq:ProofPropDetector(2,1)SlowFreqFiXedTime_First}
\Lambda_{{(2,1)},\text{p}} (\mathbf{V}) = \mathbf {\overline{b}}^H \mathbf {\overline{R}}\,\,\mathbf {\overline{B}}\,\,\mathbf {\overline{R}} \,\, \mathbf {\overline{b}},
\end{equation}
where vector $\mathbf {\overline{b}}$ is given in (\ref{Eq:DefineVectorb_Bar_ForCase(1,2)})
and matrix $\mathbf {\overline{R}}$ is given in (\ref{Eq:BlockDiagonalMAt_R_BAR_for(1,2)}). 

We use the eigenvalue decomposition of matrix $\mathbf {\overline{R}}\,\,\mathbf {\overline{B}}\,\,\mathbf {\overline{R}}$ in (\ref{Eq:ProofPropDetector(2,1)SlowFreqFiXedTime_First}) as $\bm {\widehat{\Phi}} \,\, \bm {\widehat{\Delta}} \,\,\bm {\widehat{\Phi}}^H$ where $\bm {\widehat{\Phi}}$ denotes a unitary matrix and $\bm {\widehat{\Delta}}$ denotes a diagonal matrix containing the eigenvalues, $\widehat{\lambda}_{j'}$. Define $\widehat{\mathbf c}$ and $\widehat{\bm{\mu}}$ by replacing $\bm {\overline{\Phi}}$ with $\bm {\widehat{\Phi}}$ in (\ref{Eq:DefineCOverlineInCase(1,2)}) and (\ref{Eq:DefineMU_Overline_InCase(1,2)}) respectively. Also let $\widehat{\mu}_{j'}$ and $\widehat{c}_{j'}$ be elements of vectors $\widehat{\bm{\mu}}$ and $ \widehat{\mathbf c}$ respectively. By substituting the eigenvalue decomposition into (\ref{Eq:ProofPropDetector(2,1)SlowFreqFiXedTime_First}) we get
\begin{equation}\label{Eq:ProofPropDetectorCase(2,1)SlowFreqFiXedTimeSecond2nd}
\Lambda_{{(2,1)},\text{p}} (\mathbf{V}) = 
(\widehat{\bm{\mu}}^H + \widehat{\mathbf c}^H )\widehat{\bm{\Delta}}(\bm{\widehat\mu} + \mathbf {\widehat{c}} )=\sum_{j'=1}^{KTN_R}\widehat{\lambda}_{j'}  |\widehat{\mu}_{j'}+\widehat{c}_{j'}|^2.
\end{equation}
It is observed that the right hand side of the expression in (\ref{Eq:ProofPropDetectorCase(2,1)SlowFreqFiXedTimeSecond2nd}) is a linear combination of $KTN_R$ non-central chi-squares with 2 DoFs and non-centrality parameter $\operatorname{Re}\{\widehat{\mu}_{j'}\}^2+\operatorname{Im}\{\widehat{\mu}_{j'}\}^2$, given in (\ref{Eq:InTheoremForPDforCase(2,1)SlowFreqFiXedTime}). For the null hypothesis, (\ref{Eq:InTheoremForP_FAfor(2,1)SlowFreqFiXedTime}) can be similarly proved by replacing $\mathbf v_{k,t}$ in (\ref{Eq:PropDetectorCase(2,1)MatrixFormSlowFreqFixedTime}) with $\mathbf{R}_k^\frac{1}{2}\tilde{\mathbf{v}}_{k,t}$. $\mathscr{g}$
\end{proof}

\subsection{Proposed Detector for Case (2,2): Slowly varying in Both Domains}\label{Sec:Subsection_BothSlowlyVarying}

We make use of slow variation in both the frequency and time domains and propose a detector using a rectangular two-dimensional (2-D) window. The window moves in both frequency and time. Therefore we propose to use an estimate of $\bm \alpha_{k,t}$ as $\bm \alpha_k^{(2,2)}$ that is a weighted average of a 2-D window consisting of the received voltage vectors as

\begin{equation}\label{Eq:windowingFormulaCase(2,2)SlowFreqSlowTime}
\bm {\alpha}_{k,t}^{(2,2)}=\sum_{m=-M}^M \sum_{q=-Q}^Q \mathbf{v}_{k-m,t-q} w_{k,m,t,q}.   
\end{equation}
% Conditions
% \begin{equation}\label{Eq:Weights(2,2)_casesForWin}
% \begin{split}
% &w_{k\ell}=\\
% &\begin{cases}
% 1/[(2M+1)(2Q+1)] \hspace{0.7cm}\,\,k\geq M+1, k\leq K-M,\\
% \hspace{4cm} \,\,t\geq Q+1, t\leq T-Q\\
% 1/(k+L) &k\leq L \\
% 1/(K-k+1+L) &k\geq K-L+1,
% \end{cases}
% \end{split}
% \end{equation}
with $w_{k,m,t,q}$ being the weights of the window. Parameters $M$ and $Q$ control the length of the window in the frequency domain, i.e., $2M+1$ , and in the time domain, i.e., $2Q+1$ respectively. Also, in the weighted average expression in (\ref{Eq:windowingFormulaCase(2,2)SlowFreqSlowTime}), the center of the 2-D window with $m=0$ and $q=0$ coincides with the received voltage vector for the pair $(k,t)$. Moreover, frequency and time indices are in the ranges $1$ to $K$ and $1$ to $T$ respectively. Thus, to avoid edge effects, $w_{k,m,t,q}$ is set to zero for $k-m<1$ or $k-m>K$ or $t-q<1$ or $t-q>T$. Also $w_{k,m,t,q}$ is set to zero for $m<-M$ or $m>M$ or $q<-Q$ or $q>Q$.

% vectors $\mathbf{v}_{k-m,t-q}$ exist providing that $1\leq k-m \leq K$ and $1\leq t-q \leq T$. 

There are 9 possible conditions on the weights depending on where the 2-D window lies with respect to the frequency and time domains, including edge cases which were elaborated on in Section \ref{Subsec:PropForCase(1,2)}. The weights are calculated by uniform averaging in both time and frequency and sum to one. Calculation of the weights is straightforward and similar to that of previous subsections. Hence, we omit it for brevity.

Referring to (\ref{Eq:OurNewTestStatisticMovingAVE1}) and (\ref{Eq:windowingFormulaCase(2,2)SlowFreqSlowTime}), our proposed detector for Case (2,2) is written as 
\begin{equation}\label{Eq:PropDetForCase(2,2)SlowFreqSlowTime}
\begin{split}
&\Lambda_{{(2,2)},\text{p}} (\mathbf{V}) =\\
& \sum_{k} \sum_{t} \Bigg[ \mathbf v_{k,t}^H \mathbf{R}^{-1}_k \bigg(\sum_{m} \sum_{q} \mathbf{v}_{{k-m},{t-q}} w_{k,m,t,q}\bigg) \\
&+\bigg(\sum_{m} \sum_{q} \mathbf{v}_{{k-m},{t-q}}^H w_{k,m,t,q}\bigg) \mathbf{R}^{-1}_k \mathbf{v}_{k,t}\\ & - \bigg(\sum_{m} \sum_{q} \mathbf{v}_{{k-m},{t-q}}^H w_{k,m,t,q}\bigg) \mathbf{R}^{-1}_k \\
& \hspace{2cm} \times \bigg(\sum_{m'} \sum_{q'} \mathbf{v}_{{k-m'},{t-q'}} w_{k,m',t,q'}\bigg) \Bigg].
\end{split}
\end{equation}
The detector in \eqref{Eq:PropDetForCase(2,2)SlowFreqSlowTime} can be written as a quadratic from: 
\begin{equation}\label{Eq:PropDetectorForCase(2,2)SlowFreqSlowTimeQuadratic}
\Lambda_{{(2,2)},\text{p}} (\mathbf{V}) \!=\!\sum_{k_1=1}^K\!\sum_{t_1=1}^T\!\sum_{k_2=1}^K\!\sum_{t_2=1}^T \mathbf{v}_{k_1,t_1}^H \mathbf C_{k_1t_1k_2t_2}  \mathbf{v}_{k_2,t_2},
\end{equation}
where we obtain matrix $\mathbf C_{k_1t_1k_2t_2}$ as
\begin{equation}\label{Eq:MatrixCk1t1k2t2ForCAse(2,2)SlowFreqSlowTime}
\begin{split}
\mathbf C_{k_1t_1k_2t_2} =& \mathbf{R}^{-1}_{k_1} w_{k_1,{k_1-k_2},t_1,{t_1-t_2}}  + w_{k_2,{k_2-k_1},t_2,{t_2-t_1}} \mathbf{R}^{-1}_{k_2} 
\\& - \sum_{k} \sum_{t} w_{k,{k-k_1},t,{t-t_1}} \mathbf{R}^{-1}_k w_{k,{k-k_2},t,{t-t_2}}.
\end{split}
\end{equation}
% where since $-M\leq m \leq M$ and $-Q\leq q \leq Q$, the first term exists when $-M\leq k_1-k_2 \leq M$ and $-Q\leq t_1-t_2 \leq Q$. In addition, the second term exists if $-M\leq k_2-k_1 \leq M$ and $-Q\leq t_2-t_1 \leq Q$. Finally, the third term exists when $-M\leq k-k_1 \leq M$, $-Q\leq t-t_1 \leq Q$, $-M\leq k-k_2 \leq M$ and $-Q\leq t-t_2 \leq Q$. Otherwise any term that does not satisfy its corresponding condition is set to a zero matrix.

Define a $KTN_R$ by $KTN_R$ block matrix $\mathbf{\overline{C}}$ where block $(i,j)$ is $\mathbf C_{k_1t_1k_2t_2}$ and the indices $k_1$, $t_1$, $k_2$ and $t_2$ are obtained using (\ref{Eq:IndicesInMatrixA_Bar_k1t1k2t2}), with $i=1,2,...,KT$ and $j=1,2,...,KT$. As such, referring to (\ref{Eq:PropDetectorForCase(2,2)SlowFreqSlowTimeQuadratic}), our proposed detector is expressed as 
\begin{equation}\label{Eq:PropDetectorCase(2,2)MatrixFormSlowFreq_SLOWTime}
\Lambda_{{(2,2)},\text{p}} (\mathbf{V}) = \mathbf {\overline{v}}^H \mathbf {\overline{C}}\,\, \mathbf {\overline{v}},
\end{equation}
where $\mathbf {\overline{v}}$ is given in (\ref{Eq:DefineMatrixV_Bar_ForCase(1,2)}).

In the following theorem we present an analysis for our proposed detector for Case (2,2).

\begin{theorem}\label{Theorem:TheoremForCase(2,2)SlowFreqSLOWTime}
The detector $\Lambda_{(2,2),\text{p}}(\mathbf{V})$ in (\ref{Eq:PropDetectorCase(2,2)MatrixFormSlowFreq_SLOWTime}) under $\mathcal{H}_1$ and $\mathcal{H}_0$ is respectively a linear combination of non-central chi-squares and a linear combination of chi-squares, as
\begin{equation}\label{Eq:InTheoremForPDforCase(2,2)SlowFREQSlowTime}
\Lambda_{(2,2),\text{p}}|\mathcal{H}_1=  \sum_{j'=1}^{KTN_R} {\omega}_{j'} \chi^2_{2,j'}\bigg( \operatorname{Re}\{{\zeta}_{j'}\}^2+\operatorname{Im}\{{\zeta}_{j'}\}^2\bigg),
\end{equation}
\begin{equation}\label{Eq:InTheoremForP_FAfor(2,2)SlowFREQSlowTime}
\Lambda_{(2,2),\text{p}}|\mathcal{H}_0= \sum_{j'=1}^{KTN_R} {\omega}_{j'} \chi^2_{2,j'},
\end{equation}
where ${\omega}_{j'}$ and ${\zeta}_{j'}$ are defined above \eqref{Eq:ProofPropDetectorCase(2,2)SlowSLOWTimeSecond2nd}.
\end{theorem}
\begin{proof}
Similar to the proofs of Theorems \ref{Theorem:TheoremForCase(2,1)ConstantFreqSlowTtime} and \ref{Theorem:TheoremForCase(2,1)SlowFreqFiXedTime}, (\ref{Eq:PropDetectorCase(2,2)MatrixFormSlowFreq_SLOWTime}) can be expressed as
\begin{equation}\label{Eq:ProofPropDetector(2,2)SlowFreqSLOWTime_First}
\Lambda_{{(2,2)},\text{p}} (\mathbf{V}) = \mathbf {\overline{b}}^H \mathbf {\overline{R}}\,\,\mathbf {\overline{C}}\,\,\mathbf {\overline{R}} \,\, \mathbf {\overline{b}},
\end{equation}
where $\mathbf {\overline{b}}$ and $\mathbf {\overline{R}}$ are given in (\ref{Eq:DefineVectorb_Bar_ForCase(1,2)})
and (\ref{Eq:BlockDiagonalMAt_R_BAR_for(1,2)}). Matrix $\mathbf {\overline{R}}\,\,\mathbf {\overline{C}}\,\,\mathbf {\overline{R}}$ in (\ref{Eq:ProofPropDetector(2,2)SlowFreqSLOWTime_First}) can be replaced by its eigenvalue decomposition as $\bm {\Psi} \,\, \bm {\Omega} \,\,\bm {\Psi}^H$ where $\bm {\Psi}$ denotes a unitary matrix and $\bm {\Omega}$ denotes a diagonal matrix containing the eigenvalues, ${\omega}_{j'}$. Define $\mathbf d$ and $\bm{\zeta}$ by replacing $\bm {\overline{\Phi}}$ with $\bm {\Psi}$ in (\ref{Eq:DefineCOverlineInCase(1,2)}) and (\ref{Eq:DefineMU_Overline_InCase(1,2)}) respectively. In addition, denote the elements of vectors $\bm{\zeta}$ and $ \mathbf d$ by $\zeta_{j'}$ and $d_{j'}$ respectively. Thus we obtain 
\begin{equation}\label{Eq:ProofPropDetectorCase(2,2)SlowSLOWTimeSecond2nd}
\Lambda_{{(2,2)},\text{p}} (\mathbf{V}) \!= \!
({\bm{\zeta}}^H + {\mathbf d}^H ){\bm{\Omega}}(\bm{\zeta} + \mathbf {d} )\!=\!\!\sum_{j'=1}^{KTN_R}\!\!{\omega}_{j'}  |{\zeta}_{j'}+{d}_{j'}|^2\!.
\end{equation}
Consequently the proof of (\ref{Eq:InTheoremForPDforCase(2,2)SlowFREQSlowTime}) is complete. We also refer the reader to the proofs of the previous theorems to prove (\ref{Eq:InTheoremForP_FAfor(2,2)SlowFREQSlowTime}). 

\end{proof}

\subsection{Proposed Detector for Case {(2,3)}: Slowly Varying in Frequency and Rapidly Varying in Time}\label{Subec:ProposeDetectorCase(2,3)SlowFRapidT}

% The slow variation across frequency in Case (2,3) could be owing to a correlation structure in the frequency domain. This case is relevant to the tightly coupled antenna arrays with wideband behavior that causes correlation in nearby frequency bins and consequently leads to slow variations of signal across frequency. 

We design a detector that uses a weighted average of received voltage vectors in the frequency domain as an estimate of the signal. We estimate $\bm\alpha_{k,t}$ with $\bm {\alpha}_{k,t}^{(2,3)}$ which is, for a frequency sample $k$, a weighted average of vectors $\mathbf{v}_{k,t}$ in the frequency domain as 
\begin{equation}\label{Eq:windowingFormulaMA1}
\bm {\alpha}_{k,t}^{(2,3)}=\sum_{{\ell_3}=-{L_3}}^{L_3} \mathbf{v}_{{k-\ell_3},t} w_{k,\ell_3},   
\end{equation}
where $w_{k,\ell_3}$ denote the weights. Replacing $\bm \alpha_{k,t}$ by the weighted average in (\ref{Eq:windowingFormulaMA1}) implements smoothing via a window of surrounding frequency samples. The center of the window corresponding to $\ell_3=0$ matches to the $k$-th frequency sample. Parameter $L_3$ determines the window length, i.e., $2L_3+1$, where $2L_3+1<K$. Weights $w_{k,\ell_3}$ are set to zero when $\ell_3<-L_3$ or $\ell_3>L_3$. Also, $w_{k,\ell_3}$ are set to zero for $k-\ell_3<1$ or $k-\ell_3>K$ due to edge effects.
% When $k$ and $\ell_3$ satisfy $1\leq k-\ell_3 \leq K$, vector $\mathbf{v}_{(k-\ell_3),t}$ exists and due to this when $k$ is small or large, some of the weights $w_{k,\ell_3}$ cannot have a value as their corresponding vectors $\mathbf{v}_{(k-\ell_3),t}$ do not exist. 
As explained above in Section \ref{Subsec:PropForCase(1,2)}, we obtain three conditions for the weights as
% in \cite{DyaynSlidingWin} a combined decision is made from individual detection decisions for each window displacement, no detection performance anlysis provided
\begin{equation}\label{Eq:Three_casesForWin}
w_{k,\ell_3}=
\begin{cases}
1/(2L_3+1) &k\geq L_3+1, k\leq K-L_3\\
1/(k+L_3) &k\leq L_3 \\
1/(K-k+1+L_3) &k\geq K-L_3+1,
\end{cases}
\end{equation}
which we calculate based on uniform averaging across frequency. Referring to (\ref{Eq:OurNewTestStatisticMovingAVE1}) and (\ref{Eq:windowingFormulaMA1}) we write our proposed detector for Case (2,3) as
\begin{equation}\label{Eq:DetectorForCase(2,3)FirstEq}
\begin{split}
&\Lambda_{{(2,3)},\text{p}} (\mathbf{V}) =\sum_{t} \Bigg[\sum_{k} \sum_{\ell_3=-{L_3}}^{L_3} \mathbf v_{k,t}^H \mathbf{R}^{-1}_k  \mathbf{v}_{k-\ell_3,t} w_{k,\ell_3} \\
&+ \sum_{k} \Bigg(\sum_{\ell_3=-L}^{L_3} w_{k, \ell_3} \mathbf{v}_{k-\ell_3,t}^H\Bigg) \mathbf{R}^{-1}_k \mathbf v_{k,t} 
\\ & -\!\!\! \sum_{k} \!\!\Bigg(\!\sum_{\ell_3=-L_3}^{L_3} \!\!\!\!\! w_{k,\ell_3} \mathbf{v}_{k-\ell_3,t}^H\Bigg) \! \mathbf{R}^{-1}_k \! \Bigg(\!\sum_{{\ell'_3}=-{L_3}}^{L_3} \!\!\!\!\! \mathbf{v}_{k-{\ell'_3},t} w_{k,{\ell'_3}} \Bigg) \Bigg].
\end{split}
\end{equation}
We then rewrite $\Lambda_{{(2,3)},\text{p}} (\mathbf{V})$ in (\ref{Eq:DetectorForCase(2,3)FirstEq}) in a quadratic form as 
\begin{equation}\label{Eq:logLikelihoodMatrixFormMAD1}
\Lambda_{{(2,3)},\text{p}} (\mathbf{V}) \!= \!\!\sum_{t} \!\!
\begin{bmatrix}
\mathbf v_{1,t}^H \, \mathbf v_{2,t}^H \, ... \, \mathbf v_{K,t}^H 
\end{bmatrix}
\!\! \mathbf{A} \!\!
\begin{bmatrix}
\mathbf v^T_{1,t} \, \mathbf v^T_{2,t} ... \, \mathbf v^T_{K,t}\!
\end{bmatrix}^T\!\!,
\end{equation}
where matrix $\mathbf{A}_{KN_R\times KN_R}$ is a block matrix with $N_R$ by $N_R$
% \begin{equation}\label{Eq:DefineBigMatrixA1}
% \mathbf{A}=
% % \begingroup % keep the change local
% % \setlength\arraycolsep{0.5pt}
% \begin{bmatrix}
% \mathbf{A}_{11} & \mathbf{A}_{12} & ... & \mathbf{A}_{1K}\\
% \mathbf{A}_{21} & \mathbf{A}_{21} & ... & \mathbf{A}_{2K}\\
% \vdots & \vdots & \vdots & \vdots \\
% \mathbf{A}_{K1} & \mathbf{A}_{K2} & ... & \mathbf{A}_{KK}
% \end{bmatrix},
% \end{equation}
sub-matrices $\mathbf{A}_{rs}$, for $1 \leq r \leq K$ and $1 \leq s \leq K$. We calculate the sub-matrices as 
\begin{equation}\label{Eq:ExpressionOfArsForCase(2,3)}
\begin{split}
\mathbf{A}_{rs}\!= \!\!\mathbf{R}^{-1}_r w_{r,r-s}\!\! +\! w_{s,s-r} \mathbf{R}^{-1}_s
\!\!-\! \!\sum_{k} \!\!w_{k,k-r} \mathbf{R}^{-1}_k w_{k,k-s}.
\end{split}
\end{equation}
% As $-L_3\leq \ell_3 \leq L_3$, in (\ref{Eq:ExpressionOfArsForCase(2,3)}), the first term exists when $-L_3\leq r-s \leq L_3$, the second term exists if $-L_3\leq s-r \leq L_3$ and the third term exists when $-L_3\leq k-r \leq L_3$ and $-L_3\leq k-s \leq L_3$, otherwise any term that does not satisfy its corresponding condition is replaced by a zero matrix.

We now proceed to provide a performance analysis for our proposed detector with the following theorem.
% Note that a generalized chi-square variable is an addition of quadratic, linear and constant terms.
\begin{theorem}\label{Theorem:TheoremForMAD1}
The detector $\Lambda_{(2,3),\text{p}}(\mathbf{V})$ in (\ref{Eq:logLikelihoodMatrixFormMAD1}) under $\mathcal{H}_1$ and $\mathcal{H}_0$ is respectively a linear combination of non-central chi-squares and chi-squares, as  
\begin{equation}\label{Eq:InTheoremMAD1ForPD}
\Lambda_{(2,3),\text{p}}|\mathcal{H}_1= \sum_{i=1}^{KN_R} \lambda_i \,\, \chi^2_{2T,i}\bigg(\sum_{t} \operatorname{Re}\{\mu_{i,t}\}^2+\operatorname{Im}\{\mu_{i,t}\}^2\bigg),
\end{equation}
\begin{equation}\label{Eq:InTheoremMAD1ForP_FA}
\Lambda_{(2,3),\text{p}}|\mathcal{H}_0= \sum_{i=1}^{KN_R} \lambda_i \,\, \chi^2_{2T,i},
\end{equation} 
where $\lambda_i$ and $\mu_{i,t}$ are defined above and below \eqref{Eq:mu_t_ForEq48} respectively.
\end{theorem}
\begin{proof}
Similar to the previous proofs, under $\mathcal{H}_1$ we replace $\mathbf v_{k,t}$ in (\ref{Eq:logLikelihoodMatrixFormMAD1}) with $\bm{\alpha}_{k,t}+\mathbf{R}_k^\frac{1}{2}\tilde{\mathbf{v}}_{k,t}$. Let $\tilde{\mathbf{b}}_{k,t}=\mathbf{R}_k^{-\frac{1}{2}}\bm{\alpha}_{k,t}+\tilde{\mathbf{v}}_{k,t}$ and define $\widetilde{\mathbf{R}}_{KN_R\times KN_R}$ as a block diagonal matrix with sub-matrices $\mathbf{R}_1^\frac{1}{2}$, $\mathbf{R}_2^\frac{1}{2}$, ..., $\mathbf{R}_K^\frac{1}{2}$ on the diagonal. Thus, we write $\Lambda_{(2,3),\text{p}} (\mathbf{V})$ in (\ref{Eq:logLikelihoodMatrixFormMAD1}) under $\mathcal{H}_1$ as
\begin{equation}\label{Eq:DetectorH1forPD_FactroRs1}
\Lambda_{(2,3),\text{p}}|\mathcal{H}_1 \!\! = \!\! \sum_{t} \!\!
\begin{bmatrix}
 \tilde{\mathbf{b}}_{1,t}^H \, \tilde{\mathbf{b}}_{2,t}^H \, ... \, \tilde{\mathbf{b}}_{K,t}^H 
\end{bmatrix}
\!\! \widetilde{\mathbf{R}}
\mathbf{A}
\widetilde{\mathbf{R}}
\!\!\begin{bmatrix}
\tilde{\mathbf{b}}^T_{1,t} \, \tilde{\mathbf{b}}^T_{2,t} ... \, \tilde{\mathbf{b}}^T_{K,t}
\end{bmatrix}^T.
\end{equation}
We replace matrix $\widetilde{\mathbf{R}}\mathbf{A}\widetilde{\mathbf{R}}$ in (\ref{Eq:DetectorH1forPD_FactroRs1}) by its eigenvalue decomposition $\bm{\Phi}\bm{\Delta}\bm{\Phi}^H$, where $\bm{\Phi}$ is a unitary matrix and $\bm{\Delta}$ is a diagonal matrix containing the eigenvalues $\lambda_i$ for $i=1,2,...,KN_R$.
% \begin{equation}\label{Eq:defindBlockMatrixRtild}
% \widetilde{\mathbf{R}}=
% \begin{bmatrix}
%     \mathbf{R}_1^\frac{1}{2} & \dots & \mathbf{0} \\
%     \vdots & \ddots & \vdots \\
%     \mathbf{0} & \dots & \mathbf{R}_K^\frac{1}{2}
%     \end{bmatrix}.
% \end{equation}
Define 
\begin{equation}\label{Eq:mu_t_ForEq48}
\bm{\mu}_t \! = \! \bm{\Phi}^H \!\! \begin{bmatrix} \!
(\mathbf{R}_1^{-\frac{1}{2}} \bm{\alpha}_{1,t})^T \,~ (\mathbf{R}_2^{-\frac{1}{2}} \bm{\alpha}_{2,t})^T ... \, (\mathbf{R}_K^{-\frac{1}{2}} \bm{\alpha}_{K,t})^{T} \!\!
\end{bmatrix}^T\!\!,
\end{equation}
and $\mathbf c_t=\bm{\Phi}^H \begin{bmatrix}
\tilde{\mathbf{v}}_{1,t}^T \, \tilde{\mathbf{v}}_{2,t}^T ... \, \tilde{\mathbf{v}}_{K,t}^T
\end{bmatrix}^T$. Note that $\mathbf c_t \sim \mathcal{CN}(\mathbf 0,\, \mathbf I)$ as $\bm{\Phi}$ is  unitary. Also denote by $\mu_{i,t}$ and $c_{i,t}$ the $i$-th elements of $\bm \mu_t$ and $\mathbf{c}_t$ respectively. Thus, (\ref{Eq:DetectorH1forPD_FactroRs1}) is given as
\begin{equation}\label{Eq:muDeltaCforProposed1}
\Lambda_{(2,3),\text{p}}|\mathcal{H}_1 \!\! =\!\! \sum_{t} \! (\bm{\mu}_t^H \!+\! \mathbf c_t^H \!)\bm{\Delta}(\bm{\mu}_t \!+\! \mathbf c_t \!)\!=\!\!\!\sum_{i=1}^{KN_R}\!\!\lambda_i \!\! \sum_{t} |\mu_{i,t}\!+\!c_{i,t}|^2.
\end{equation}

Finally, the summation over $t$ in (\ref{Eq:muDeltaCforProposed1}) is a chi-square with $2T$ DoFs and non-centrality parameter $\sum_t (\operatorname{Re}\{\mu_{i,t}\}^2+\operatorname{Im}\{\mu_{i,t}\}^2)$. Hence, (\ref{Eq:muDeltaCforProposed1}) is a linear combination of non-central chi-squares, which results in (\ref{Eq:InTheoremMAD1ForPD}). Note that for the null hypothesis there is no signal component. Thus, to calculate $\Lambda_{(2,3),\text{p}}|\mathcal{H}_0$ we replace $\mathbf v_{k,t}$ in (\ref{Eq:logLikelihoodMatrixFormMAD1}) with $\mathbf{R}_k^\frac{1}{2}\tilde{\mathbf{v}}_{k,t}$ and repeat the same process, which gives the expression in (\ref{Eq:InTheoremMAD1ForP_FA}).
\end{proof}

% \begin{equation}
% P_{FA,P}=\int_{\gamma_1}^\infty p(\Lambda|\mathcal{H}_0) d \Lambda
% \end{equation}

\subsection{Proposed Detector for Case (3,2): Rapidly Varying in Frequency and Slowly Varying in Time}\label{Sec:Case(3,2)}

For Case (3,2), considering a slowly varying signal in time we propose a detector using a weighted average of received voltage vectors in the time domain. We estimate $\bm \alpha_{k,t}$ by $\bm {\alpha}_{k,t}^{(3,2)}$ which is, for time sample $t$, a weighted average taken in the time domain as
\begin{equation}\label{Eq:windowingFormulaCase(3,2)}
\bm\alpha_{k,t}^{(3,2)}=\sum_{{\ell_4}=-{L_4}}^{L_4} \mathbf{v}_{k,t-{\ell_4}} w_{t,{\ell_4}},   
\end{equation}
where $w_{t,\ell_4}$ denotes the weights. We use (\ref{Eq:windowingFormulaCase(3,2)}) to replace $\bm \alpha_{k,t}$ by a window in the time domain whose center, corresponding to $\ell_4=0$, matches the $t$-th time sample. The length of the window is $2L_4+1$ which is assumed to be smaller than T. To avoid edge cases, we set weights $w_{t,\ell_4}$ to zero when $t-\ell_4<1$ or $t-\ell_4>T$. Also $w_{t,\ell_4}=0$ for $\ell_4<-L_4$ or $\ell_4>L_4$.
% , and the larger $L_4$ gets the better detection performance we get due to more averaging.
% Vector $\mathbf{v}_{k,t-\ell_4}$ exists for $1\leq t-\ell_4 \leq T$ with corresponding weights $w_{t,\ell_4}$. 
We calculate the weights by uniform averaging in time as
\begin{equation}\label{Eq:Three_casesForWinCase(3,2)}
w_{t,\ell_4}=
\begin{cases}
1/(2L_4+1) &t\geq L_4+1, t\leq T-L_4\\
1/(t+L_4) &t\leq L_4 \\
1/(T-t+1+L_4) &t\geq T-L_4+1.
\end{cases}
\end{equation}

We write our proposed detector for Case (3,2), referring to \eqref{Eq:OurNewTestStatisticMovingAVE1} and (\ref{Eq:windowingFormulaCase(3,2)}), as
\begin{equation}\label{Eq:DetectorForCase(3,2)FirstEquation}
\begin{split}
&\Lambda_{{(3,2)},\text{p}} (\mathbf{V}) =\!\!\sum_k \!\!\Bigg[\!\!\sum_t \mathbf v_{k,t}^H \mathbf{R}^{-1}_k\! \Bigg(\!\sum_{{\ell_4}=-{L_4}}^{L_4}   \mathbf{v}_{k,t-{\ell_4}} w_{t,{\ell_4}}\!\!\Bigg) \\
&+ \sum_t \Bigg(\sum_{{\ell_4}=-{L_4}}^{L_4} w_{t,{\ell_4}} \mathbf{v}_{k,t-{\ell_4}}^H\Bigg) \mathbf{R}^{-1}_k \mathbf v_{k,t} 
\\ & - \!\! \sum_t\!\! \Bigg(\!\sum_{{\ell_4}=-{L_4}}^{L_4}\!\! w_{t,{\ell_4}} \mathbf{v}_{k,t-{\ell_4}}^H\!\!\Bigg) \!\mathbf{R}^{-1}_k\! \Bigg(\!\sum_{ {\ell'_4}=-{L_4} }^{L_4} \!\mathbf{v}_{k,t-{\ell'_4}} w_{t,{\ell'_4}} \!\!\Bigg) \!\Bigg]\!.
\end{split}
\end{equation}
We then rewrite $\Lambda_{{(3,2)},\text{p}} (\mathbf{V})$ in (\ref{Eq:DetectorForCase(3,2)FirstEquation}) in a quadratic form as
\begin{equation}\label{Eq:logLikelihoodMatrixFormMACase(3,2)}
\Lambda_{{(3,2)},\text{p}} (\!\mathbf{V}\!) \!\!= \! \!\sum_k \!\!
\begin{bmatrix}
\!\mathbf v_{k,1}^H \, \mathbf v_{k,2}^H \, ...  \mathbf v_{k,T}^H \!
\end{bmatrix}
\!\! \mathbf{A}_k \!\!
\begin{bmatrix}
\! \mathbf v^T_{k,1} \, \mathbf v^T_{k,2} ...  \mathbf v^T_{k,T}
\! \end{bmatrix}^T\!\!,
\end{equation}
where $\mathbf{A}_k$ is a $TN_R$ by $TN_R$ block matrix whose block $(r',s')$ is an $N_R$ by $N_R$ matrix denoted by $\mathbf{A}_{r's',k}$ for $1 \leq r' \leq T$ and $1 \leq s' \leq T$. We calculate the blocks as 
\begin{equation}\label{Eq:expressionOfArdashSdashAndkCase(3,2)}
\mathbf{A}_{r's',k}\!=  z_{r's'} \mathbf{R}^{-1}_k,
\end{equation}
where the scalar $z_{r's'}$ is defined as
\begin{equation}\label{Eq:DefineZrDashSdashCase(3,2)}
z_{r's'}=w_{r',r'-s'\!}\! +\! w_{s',s'-r'\!}
-\! \!\sum_t \!w_{t,t-r'} w_{t,t-s'\!}.
\end{equation}
% As $-L_4\leq \ell_4 \leq L_4$, in (\ref{Eq:ExpressionOfArsForCase(2,3)}), the first term exists when $-L_4\leq r'-s' \leq L_4$, the second term exists if $-L_4\leq s'-r' \leq L_4$ and the third term exists when $-L_4\leq t-r' \leq L_4$ and $-L_4\leq t-s' \leq L_4$, otherwise any coefficient not satisfying its corresponding condition is set to zero.

We now formalize our performance analysis for our proposed detector, $\Lambda_{{(3,2)},p}$, in the following theorem.

\begin{theorem}\label{Theorem:TheoremForMA_Case(3,2)RapidF_SlowT}
The detector $\Lambda_{(3,2),\text{p}}(\mathbf{V})$ in (\ref{Eq:logLikelihoodMatrixFormMACase(3,2)}) under $\mathcal{H}_1$ and $\mathcal{H}_0$ is respectively a linear combination of non-central chi-squares and chi-squares, as
\begin{equation}\label{Eq:InTheoremForPD_MA_forCase(3,2)}
\Lambda_{(3,2),\text{p}}|\mathcal{H}_1=  \sum_{i'=1}^{TN_R} \lambda_{i'} \,\, \chi^2_{2K,i'}\bigg( \sum_k \operatorname{Re}\{\mu_{i',k}\}^2+\operatorname{Im}\{\mu_{i',k}\}^2\bigg),
\end{equation}
\begin{equation}\label{Eq:InTheoremForP_FA_MA_forCase(3,2)}
\Lambda_{(3,2),\text{p}}|\mathcal{H}_0= \sum_{i'=1}^{TN_R} \lambda_{i'} \chi^2_{2K,i'},
\end{equation}
where $\lambda_{i'}$ and $\mu_{i',k}$ are defined above and below \eqref{Eq:DefineMuKVecWithPhiForTheorem5} respectively.
\end{theorem}
\begin{proof}
Define $\tilde{\mathbf{b}}_{k,t}=\mathbf{R}_k^{-\frac{1}{2}}\bm{\alpha}_{k,t}+\tilde{\mathbf{v}}_{k,t}$ and matrix $\widetilde{\mathbf G}$ as $\widetilde{\mathbf G}=\widetilde{\mathbf{R}}_k\mathbf{A}_k\widetilde{\mathbf{R}}_k$ with $\widetilde{\mathbf{R}}_k$ being the $TN_R$ by $TN_R$ block diagonal matrix, $\widetilde{\mathbf{R}}_k=\text{blockdiag}\{\mathbf{R}_k^\frac{1}{2},...,\mathbf{R}_k^\frac{1}{2}\}$. As in previous proofs, we substitute $\bm{\alpha}_{k,t}+\mathbf{R}_k^\frac{1}{2}\tilde{\mathbf{v}}_{k,t}$ for $\mathbf{v}_{k,t}$ in (\ref{Eq:logLikelihoodMatrixFormMACase(3,2)}) to obtain $\Lambda_{(3,2),\text{p}}|\mathcal{H}_1$ as
\begin{equation}\label{Eq:DetectorH1forPDCase(3,2)_Factor_Rks}
\Lambda_{(3,2),\text{p}}|\mathcal{H}_1= \!\! \sum_k \!\!
\begin{bmatrix}
\tilde{\mathbf{b}}_{k,1}^H \, \tilde{\mathbf{b}}_{k,2}^H \, ... \, \tilde{\mathbf{b}}_{k,T}^H 
\end{bmatrix}
\widetilde{\mathbf G}
\!\!\begin{bmatrix}
\tilde{\mathbf{b}}^T_{k,1} \, \tilde{\mathbf{b}}^T_{k,2} ... \, \tilde{\mathbf{b}}^T_{k,T}
\end{bmatrix}^T.
\end{equation}
Matrix $\widetilde{\mathbf G}$ does not depend on $k$ since, referring to (\ref{Eq:expressionOfArdashSdashAndkCase(3,2)}), it can be simplified as $\widetilde{\mathbf G}= \mathbf{Z}' \otimes \mathbf{I}_{N_R}$, where $\mathbf{Z}'$ denote a $T$ by $T$ matrix including $z_{r's'}$. Also $\otimes$ indicates the Kronecker product. We replace matrix $\widetilde{\mathbf{G}}$ in (\ref{Eq:DetectorH1forPDCase(3,2)_Factor_Rks}) by its eigenvalue decomposition $\widetilde{\bm{\Phi}} \widetilde{\bm{\Delta}}  \widetilde{\bm{\Phi}}^H$, where $\widetilde{\bm{\Phi}}$ is a unitary matrix and $\widetilde{\bm{\Delta}}$ is a diagonal matrix containing the eigenvalues, $\lambda_{i'}$, where $i'=1,2,...,TN_R$.
% \begin{pmatrix}
% z_{11}\mathbf{I}_{N_R} & z_{12}\mathbf{I}_{N_R} & ...& z_{1T}\mathbf{I}_{N_R}\\
% z_{21}\mathbf{I}_{N_R} & z_{22}\mathbf{I}_{N_R} & ...& z_{2T}\mathbf{I}_{N_R}\\ \vdots & \vdots & \vdots & \vdots &\\
% z_{T1}\mathbf{I}_{N_R} & z_{T2}\mathbf{I}_{N_R} & ...& z_{TT}\mathbf{I}_{N_R}\\
% \end{pmatrix},
Define 
\begin{equation}\label{Eq:DefineMuKVecWithPhiForTheorem5}
\bm{\mu}_k \! = \! \widetilde{\bm{\Phi}}^H \!\! \begin{bmatrix} \!
(\mathbf{R}_k^{-\frac{1}{2}} \bm{\alpha}_{k,1})^T \,~ (\mathbf{R}_k^{-\frac{1}{2}} \bm{\alpha}_{k,2})^T ... \, (\mathbf{R}_k^{-\frac{1}{2}} \bm{\alpha}_{k,T})^T \!
\end{bmatrix}^T,
\end{equation}
and $\mathbf c_k=\widetilde{\bm{\Phi}}^H \begin{bmatrix}
\tilde{\mathbf{v}}_{k,1}^T \, \tilde{\mathbf{v}}_{k,2}^T ... \, \tilde{\mathbf{v}}_{k,T}^T
\end{bmatrix}^T$. Note that $\mathbf c_k \sim \mathcal{CN}(\mathbf 0,\, \mathbf I)$ since $\widetilde{\bm{\Phi}}$ is a unitary matrix. Also, denote the $i'$-th elements of $\bm \mu_k$ and $\mathbf{c}_k$ by $\mu_{i',k}$ and $c_{i',k}$ respectively. Thus, (\ref{Eq:DetectorH1forPDCase(3,2)_Factor_Rks}) is given as
\begin{equation}\label{Eq:mu_kDeltaCforProposedForCase(3,2)}
\begin{split}
\Lambda&_{(3,2),\text{p}}|\mathcal{H}_1 = \\
&\sum_k  (\bm{\mu}_k^H \!+\! \mathbf c_k^H \!)\widetilde{\bm{\Delta}}(\bm{\mu}_k \!+\! \mathbf c_k )\!=\!\sum_{i'=1}^{TN_R}\lambda_{i'}\sum_k  |\mu_{i',k}\!+\!c_{i',k}|^2.
\end{split}
\end{equation}
Finally, the expression in (\ref{Eq:mu_kDeltaCforProposedForCase(3,2)}) is a linear combination of chi-squares with $2K$ DoFs and non-centrality parameter $\sum_k (\operatorname{Re}\{\mu_{i',k}\}^2+\operatorname{Im}\{\mu_{i',k}\}^2)$, resulting in (\ref{Eq:InTheoremForPD_MA_forCase(3,2)}). To calculate $\Lambda_{(3,2),\text{p}}|\mathcal{H}_0$, note that there is no signal component under null hypothesis $\mathcal{H}_0$. Thus, we replace $\mathbf v_{k,t}$ in (\ref{Eq:logLikelihoodMatrixFormMACase(3,2)}) with $\mathbf{R}_k^\frac{1}{2}\tilde{\mathbf{v}}_{k,t}$ and repeat the same process, giving the expression in (\ref{Eq:InTheoremForP_FA_MA_forCase(3,2)}).
\end{proof}

\section{Analysis for a Signal Without Slow Variations}\label{Sec:SignalWithoutSlowVar}

We now proceed to derive detectors and provide our analysis for a signal without slow variations, considering frequency-dependent noise correlation. These are the cases where the signal is constant or rapidly varying in a domain, i.e., Case (1,1), Case (1,3), Case (3,1) and Case (3,3). For these detectors, the subscript $\text{p}$ is omitted, as we derive them based on the MLE method, rather than our proposed approach for signals with slow variations.

\subsection{Analysis for Case {(1,1)}: Constant in Both Domains}\label{Sec:subSecConstantInBothD}
This is the simplest case where the signal is constant in both frequency and time domains so that we can drop the indices $k$ and $t$ from $\bm \alpha_{k,t}$, i.e., $\bm \alpha_{k,t}=\bm \alpha$. The LLR of \eqref{Eq:LRT_movingAve1} for this case is written as
\begin{equation}\label{Eq:LogLiklihoodforBothConstant}
\begin{split}
\Lambda_{(1,1)}(\mathbf{V}) &= -  \bm {\alpha}^H\left(T\sum_k \mathbf{R}^{-1}_k\right)\bm{\alpha}\\
&+2\operatorname{Re}\left\{\bm {\alpha}^H \sum_k \left(\mathbf{R}^{-1}_k \sum_t \mathbf{v}_{k,t}\right)\right\}.
\end{split}
\end{equation}
As the signal $\bm \alpha$ is unknown, referring to \eqref{Eq:logLikelihoodForGLRTwithMLEbased1}, we use the MLE of $\bm \alpha$ which is denoted by $\bm \alpha^*$. The expression in (\ref{Eq:LogLiklihoodforBothConstant}) has a quadratic structure. Thus, we compute the maximizer of (\ref{Eq:LogLiklihoodforBothConstant}) as $\bm \alpha^*=\left(T\sum_k \mathbf{R}^{-1}_k\right)^{-1}\sum_k \left(\mathbf{R}^{-1}_k \sum_t \mathbf{v}_{k,t}\right)$, giving the detector for Case (1,1) as
\begin{equation}\label{Eq:LogLiklihoodforBothConstantOptimized}
\begin{split}
\Lambda^*_{(1,1)}(\mathbf{V}) \! = \! \left(\!\sum_k \!\sum_t \mathbf{v}_{k,t}^H \mathbf{R}^{-1}_k\!\right)  \!\mathbf G \!
\left(\!\sum_k\! \mathbf{R}^{-1}_k \!\sum_t\! \mathbf{v}_{k,t}\right)\!,
\end{split}
\end{equation}
where $\mathbf G=\left(T\sum_k \mathbf{R}^{-1}_k\right)^{-1}$. Next, we provide analysis for Case (1,1). The detector for this case can be used for time and frequency varying signals, hence we conduct analysis for a general signal model indexed by both $k$ and $t$ as $\bm \alpha_{k,t}$.

As in Section \ref{Sec:ProposedForAtleastOneSlow2}, under $\mathcal{H}_1$, we replace $\mathbf v_{k,t}$ in (\ref{Eq:LogLiklihoodforBothConstantOptimized}) with $\bm{\alpha}_{k,t}+\mathbf{R}_k^\frac{1}{2}\tilde{\mathbf{v}}_{k,t}$ which gives
\begin{equation}\label{Eq:proofof1and1ReplaceVfrist}
\begin{split}    
\Lambda^*_{(1,1)}|\mathcal{H}_1=\Bigg[\sum_k &\left(\sum_t \bm \alpha_{k,t}^H\right) \mathbf{R}_k^{-1} + \mathbf{a}^H \Bigg] \mathbf G \\ &\left[ \sum_k \left(\mathbf{R}_k^{-1} \sum_t \bm \alpha_{k,t}\right) + \mathbf{a}\right],
\end{split}
\end{equation}
where $\mathbf{a}=\sum_k \mathbf{R}_k^{-\frac{1}{2}} \sum_t \tilde{\mathbf{v}}_{k,t}$. Note that $\mathbf{a}$ is a zero-mean Gaussian vector with covariance matrix $T\sum_k \mathbf{R}^{-1}_k$ which is equal to $\mathbf G^{-1}$. Therefore we can rewrite $\mathbf{a}$ as $\mathbf{a}=\mathbf G^{-\frac{1}{2}}\tilde{\mathbf{w}}$ where $\tilde{\mathbf{w}} \sim \mathcal{CN}(\mathbf 0,\, \mathbf I)$. Also we can factor $\mathbf G$ in (\ref{Eq:proofof1and1ReplaceVfrist}) as $\mathbf G^{\frac{1}{2}}\mathbf G^{\frac{1}{2}}$ to rewrite (\ref{Eq:proofof1and1ReplaceVfrist}) as
\begin{equation}\label{Eq:proofof1and1ReplaceVSec2}   
\Lambda^*_{(1,1)}|\mathcal{H}_1=(\hat{\mathbf x} + \tilde{\mathbf{w}})^H(\hat{\mathbf x} + \tilde{\mathbf{w}}),
\end{equation}
where $\hat{\mathbf x}= \mathbf G^{\frac{1}{2}} \sum_k (\mathbf{R}^{-1}_k \sum_t \bm \alpha_{k,t})$. As $\tilde{\mathbf{w}} \sim \mathcal{CN}(\mathbf 0,\, \mathbf I)$, the expression in (\ref{Eq:proofof1and1ReplaceVSec2}) is a non-central chi-square. By defining $\hat{x}_j$ as the $j$-th element of $\hat{\mathbf x}$, the detector $\Lambda^*_{(1,1)}(\mathbf{V})$ under $\mathcal{H}_1$ is a non-central chi-square as
\begin{equation}\label{Eq:BothConstant1and1H1}
\Lambda^*_{(1,1)}|\mathcal{H}_1= \chi^2_{2 N_R}\bigg( \sum_{j=1}^{N_R} \operatorname{Re}\{\hat{x}_j\}^2+\operatorname{Im}\{\hat{x}_j\}^2\bigg).
\end{equation} 

For the null hypothesis there is no signal component. Thus, to calculate $\Lambda^*_{(1,1)}|\mathcal{H}_0$ we replace $\mathbf v_{k,t}$ in (\ref{Eq:LogLiklihoodforBothConstantOptimized}) with $\mathbf{R}_k^\frac{1}{2}\tilde{\mathbf{v}}_{k,t}$ and repeat the above process, resulting in a chi-square as
\begin{equation}\label{Eq:BothConstant1and1FAH_0}
\Lambda^*_{(1,1)}|\mathcal{H}_0=\chi^2_{2 N_R}.
\end{equation}

\subsection{Analysis for Case {(1,3)}: Constant in Frequency and Rapidly Varying in Time}\label{Sec:ConstantDetector}

We drop $k$ from $\bm{\alpha}_{k,t}$ to have a constant signal in the frequency domain but varying in time. Define $\bm \beta = \begin{bmatrix}
\bm \alpha_1^T \, \bm \alpha_2^T \, ... \, \bm \alpha_T^T  
\end{bmatrix}^T$, $\mathbf{M}$ as a block diagonal matrix whose all diagonal blocks are $\sum_k \mathbf R^{-1}_k$ or $\mathbf{M}=\text{blockdiag}\{\sum_k \mathbf R^{-1}_k,...,\sum_k \mathbf R^{-1}_k\}$, and $\mathbf{u}=\begin{bmatrix}(\sum_k \mathbf R^{-1}_k\mathbf
v_{k,1})^T\, (\sum_k \mathbf R^{-1}_k\mathbf
v_{k,2})^T\,...\,(\sum_k \mathbf R^{-1}_k\mathbf
v_{k,T})^T\end{bmatrix}^T$. Hence, the LLR of \eqref{Eq:LRT_movingAve1} is given as
\begin{equation}
\begin{split}
\Lambda_{(1,3)} (\mathbf{V})\!\!=& 2 \!\operatorname{Re}\!\Big\{\!\sum_t \! \bm \alpha_t^H \! \sum_k \! \mathbf R^{-1}_k\mathbf v_{k,t} \!\Big\}\!-\!\!\!\sum_t \!\bm \alpha_t^H \! \Big(\!\sum_k \mathbf R^{-1}_k\!\Big) \bm \alpha_t \\= & 2\operatorname{Re}\{\bm \beta^H\mathbf{u}\}-\bm \beta^H \mathbf{M} \bm \beta.
\end{split}
\end{equation}
Since the signal is unknown we use the MLE of $\bm{\beta}$ as $\bm \beta^*=\mathbf M^{-1}\mathbf{u}$, which is equivalent to replacing the MLE of $\bm \alpha_t$ in \eqref{Eq:logLikelihoodForGLRTwithMLEbased1}. This gives the detector as 
\begin{equation}\label{Eq:LambdaStarConstantD1}
\Lambda^*_{(1,3)}(\mathbf V)\!\!=\!\!\mathbf u^H \mathbf{M}^{-1} \mathbf{u}\!=\!\!\!\sum_t \!\!\bigg(\!\!\!\sum_k \!\mathbf{v}_{k,t}^H \mathbf R^{-1}_k\!\!\bigg)\! \mathbf{S} \! \bigg(\!\!\sum_k \! \mathbf R^{-1}_k \mathbf{v}_{k,t}\!\!\bigg),
\end{equation}
where $\mathbf S=(\sum_k \mathbf R^{-1}_k)^{-1}$. We now present our analysis for Case (1,3).

The detector can be used for frequency-varying signals, thus we perform the analysis for a general signal model indexed by both $k$ and $t$ as $\bm \alpha_{k,t}$. Similar to previous analysis under $\mathcal{H}_1$, (\ref{Eq:LambdaStarConstantD1}) is rewritten as
\begin{equation}\label{Eq:InTheoremTWoforCnsDH1}
\Lambda^*_{(1,3)}|\mathcal{H}_1\!\!=\!\!\!\sum_t \!\!\bigg(\!\! \! \sum_k \bm{\alpha}_{k,t}^H \mathbf R^{-1}_k+\mathbf{f}_t^H\!\!\bigg)\! \mathbf{S} \! \bigg(\!\!\sum_k \! \mathbf R^{-1}_k \bm{\alpha}_{k,t}+\mathbf{f}_t\!\!\bigg),
\end{equation}
where $\mathbf{f}_t=\sum_k \mathbf R_k^{-\frac{1}{2}} \tilde{\mathbf{v}}_{k,t}$ and thus $\mathbf{f}_t$ is a zero-mean Gaussian vector with covariance matrix $\mathbf{S}^{-1}$. Replacing $\mathbf{f}_t$ with $\mathbf{S}^{-\frac{1}{2}}\mathbf{w}_t$ where $\mathbf{w}_t \sim \mathcal{CN}(\mathbf 0,\, \mathbf I)$ and replacing $\mathbf{S}$ in (\ref{Eq:InTheoremTWoforCnsDH1}) with $\mathbf{S}^{\frac{1}{2}}\mathbf{S}^{\frac{1}{2}}$ we rewrite (\ref{Eq:InTheoremTWoforCnsDH1}) as
\begin{equation}\label{Eq:InTheoremTWoforCnsDHArriveAtChi2}
\Lambda_{(1,3)}^*|\mathcal{H}_1 \! =\! \sum_t  (\mathbf{q}_t \!+\! \mathbf w_t )^H(\mathbf{q}_t \!+\! \mathbf w_t ),
\end{equation}
where $\mathbf{q}_t=\mathbf{S}^{\frac{1}{2}}\sum_k \! \mathbf R_k^{-1} \bm{\alpha}_{k,t}$, whose $j$-th element is $q_{j,t}$. The $t$-th term in (\ref{Eq:InTheoremTWoforCnsDHArriveAtChi2}) is a non-central chi-square $\chi^2_{2 N_R}(\sum_{j=1}^{N_R} \operatorname{Re}\{q_{j,t}\}^2+\operatorname{Im}\{q_{j,t}\}^2)$ and therefore (\ref{Eq:InTheoremTWoforCnsDHArriveAtChi2}) is an addition of non-central chi-squares which is also a non-central chi-square as
\begin{equation}\label{Eq:InTheoremConstantD1ForPD}
\Lambda_{(1,3)}^*|\mathcal{H}_1= \chi^2_{2 N_R T}\bigg(\sum_t \sum_{j=1}^{N_R} \operatorname{Re}\{q_{j,t}\}^2+\operatorname{Im}\{q_{j,t}\}^2\bigg).
\end{equation}

Following the same approach and replacing $\mathbf v_{k,t}$ in (\ref{Eq:LambdaStarConstantD1}) with $\mathbf{R}_k^\frac{1}{2}\tilde{\mathbf{v}}_{k,t}$, the detector $\Lambda^*_{(1,3)}(\mathbf V)$ under  $\mathcal{H}_0$ is also obtained, which is a chi-square as follows
\begin{equation}\label{Eq:InTheoremConstantDForPFalse}
\Lambda_{(1,3)}^*|\mathcal{H}_0= \chi^2_{2 N_R T}.
\end{equation}

\subsection{Analysis for Case {(3,1)}: Rapidly Varying in Frequency and Constant in Time}\label{Sec:DetectorCase{(3,1)}RapidFreqConsTime}

For case {(3,1)}, the time index $t$ is dropped from the signal and only the frequency index is considered, i.e, $\bm \alpha_k$. Define $\tilde{\bm \alpha} = \begin{bmatrix} \bm \alpha_1^T \, \bm \alpha_2^T \, ... \, \bm \alpha_K^T \end{bmatrix}^T$,  $\tilde{\mathbf{M}}$ as a block diagonal matrix with blocks $T \mathbf R^{-1}_1$, $T \mathbf R^{-1}_2$, ..., $T \mathbf R^{-1}_K$ on its diagonal, i.e, $\tilde{\mathbf{M}}=\text{blockdiag}\{T \mathbf R^{-1}_1$, $T \mathbf R^{-1}_2$, ..., $T \mathbf R^{-1}_K\}$, and $\tilde{\mathbf{u}}=\begin{bmatrix}
\mathbf R^{-1}_1 \sum_t \mathbf v_{1,t}, \mathbf R^{-1}_2 \sum_t \mathbf v_{2,t} \, ... \, \mathbf R^{-1}_K \sum_t \mathbf v_{K,t}
\end{bmatrix}$. Using the LLR of (\ref{Eq:LRT_movingAve1}), we write the detector for this case as
\begin{equation}\label{Eq:RapidFreqConsTimeFirstEqu1}
\begin{split}
\Lambda_{(3,1)} \!(\mathbf{V})\!\!=& 2\! \operatorname{Re}\!\Big\{\!\!\sum_k \!\!\bm \alpha_k^H \mathbf R^{-1}_k \!\!\sum_t\!\! \mathbf v_{k,t} \!\Big\}\!\!-\!T \!\sum_k\!\bm \alpha_k^H \mathbf R^{-1}_k \!\bm \alpha_k \\= & 2\operatorname{Re}\{\tilde{\bm \alpha}^H\tilde{\mathbf{u}}\}-\tilde{\bm \alpha}^H \tilde{\mathbf{M}} \tilde{\bm \alpha}.
\end{split}
\end{equation}
The signal is unknown thus we use the MLE of $\tilde{\bm \alpha}$ as $\tilde{\bm \alpha}^*=\tilde{\mathbf{M}}^{-1} \tilde{\mathbf{u}}$, which is equivalent to substituting the MLE of $\bm \alpha_k$ in \eqref{Eq:logLikelihoodForGLRTwithMLEbased1}, resulting in a detector as follows:
\begin{equation}\label{Eq:RapidFreqConsTimeFirstEqu2Opt}
\Lambda_{(3,1)}^* (\mathbf{V})\!=\!\tilde{\mathbf{u}}^H \tilde{\mathbf{M}}^{-1} \tilde{\mathbf{u}}\! = \!\!\frac{1}{T} \!\! \sum_k \!\! \left(\!\!\sum_t \mathbf{v}_{k,t}^H \mathbf R^{-1}_k \! \! \right)\!\!\!\left(\!\sum_t \mathbf{v}_{k,t}\!\! \right)\!\!.
\end{equation}
We now proceed to provide analysis for the detector for Case {(3,1)}. Note that since the detector $\Lambda_{(3,1)}^*(\mathbf{V})$ can be used in a general detection scenario where the signal is varying in time, we use both time and frequency indices for the signal in our analysis. Hence, similar to previous analysis, the detector under $\mathcal{H}_1$ is obtained as
\begin{equation}\label{Eq:InTheProofForCase3and1probDetectFirst}
\begin{split}
\Lambda_{(3,1)}^*|\mathcal{H}_1= \sum_k \bigg[&\frac{1}{\sqrt{T}}\sum_t \bm \alpha_{k,t}^H  \mathbf{R}_k^{-\frac{1}{2}} + \frac{1}{\sqrt{T}}\sum_t \tilde{\mathbf{v}}^H_{k,t} \bigg] \\ &\bigg[\frac{1}{\sqrt{T}} \mathbf{R}_k^{-\frac{1}{2}} \sum_t  \bm \alpha_{k,t}\! +\! \frac{1}{\sqrt{T}}\sum_t \tilde{\mathbf{v}}_{k,t} \bigg]\!. 
\end{split}
\end{equation}
We rewrite (\ref{Eq:InTheProofForCase3and1probDetectFirst}) as
\begin{equation}\label{Eq:InTheProofForCase3,1probDetect2nd}
\Lambda_{(3,1)}^*|\mathcal{H}_1  = \sum_k  (\bm{\ell}_k + \tilde{\bm{\ell}}_k )^H(\bm{\ell}_k + \tilde{\bm{\ell}}_k ),
\end{equation}
where ${\bm{\ell}}_k \!\!=\!\!\frac{1}{\sqrt{T}} \mathbf{R}_k^{-\frac{1}{2}} \!\!\sum_t \!\bm \alpha_{k,t}$ and $\tilde{{\bm{\ell}}}_k \!\!=\!\! \frac{1}{\sqrt{T}}\!\sum_t \!\!\tilde{\mathbf{v}}_{k,t}$. The $k$-th term in the summation in (\ref{Eq:InTheProofForCase3,1probDetect2nd}) is a non-central chi-square as $\chi^2_{2 N_R}(\sum_{j=1}^{N_R} \operatorname{Re}\{l_{j,k}\}^2+\operatorname{Im}\{l_{j,k}\}^2)$ where $l_{j,k}$ is the $j$-th component of vector ${\bm{\ell}}_k$. Therefore (\ref{Eq:InTheProofForCase3,1probDetect2nd}) is an addition of non-central chi-squares which is another non-central chi-square as
\begin{equation}\label{Eq:InTheoremForCase3and1PD}
\Lambda_{(3,1)}^*|\mathcal{H}_1= \chi^2_{2 N_R K}\bigg(\sum_k \sum_{j=1}^{N_R} \operatorname{Re}\{l_{j,k}\}^2+\operatorname{Im}\{l_{j,k}\}^2\bigg).
\end{equation}
To compute $\Lambda_{(3,1)}^*|\mathcal{H}_0$, similarly we substitute $\mathbf{R}_k^\frac{1}{2}\tilde{\mathbf{v}}_{k,t}$ for $\mathbf{v}_{k,t}$ in (\ref{Eq:RapidFreqConsTimeFirstEqu2Opt}) and repeat a similar process, giving a chi-square as 
\begin{equation}\label{Eq:InTheoremForCase3and1FlaseA}
\Lambda_{(3,1)}^*|\mathcal{H}_0= \chi^2_{2 N_R K}.
\end{equation}

\subsection{Extended Energy Detector for Case {(3,3)}: Rapidly Varying in Both Frequency and Time}\label{Sec:BaselineWideBdetector}

We derive a detector for a signal rapidly varying in both frequency and time. The detector that we derive is an extended energy detector for noise with a frequency-dependent covariance matrix as well as both frequency and time samples. Define two vectors
\begin{equation}
\begin{split}
&\widehat{\bm \alpha}\!\!=\!\!\!\begin{bmatrix}\!
\bm \alpha_{1,1}^T \, \bm \alpha_{2,1}^T \,\! ...\! \, \bm \alpha_{K,1}^T \, \bm \alpha_{1,2}^T \, \bm \alpha_{2,2}^T \,\! ... \!\, \bm \alpha_{K,2}^T\, ... \, \bm \alpha_{1,T}^T \, \bm \alpha_{2,T}^T \,\! ... \!\, \bm \alpha_{K,T}^T\!\!
\end{bmatrix}^T,\\
&\widehat{\mathbf{v}}\!=\!\!\begin{bmatrix}\!
\mathbf{v}_{1,1}^T \, \mathbf{v}_{2,1}^T \,\! ...\! \, \mathbf{v}_{K,1}^T \, \mathbf{v}_{1,2}^T \, \mathbf{v}_{2,2}^T \,\! ... \!\, \mathbf{v}_{K,2}^T\, ... \, \mathbf{v}_{1,T}^T \, \mathbf{v}_{2,T}^T \,\! ... \!\, \mathbf{v}_{K,T}^T\!\!
\end{bmatrix}^T.
% &\widehat{\mathbf{R}}\!=\!\text{diag}\!\!\begin{bmatrix}\!
% \mathbf{R}_1\!,\!\! \, \mathbf{R}_2\!,\!\! \, ...\! \, ,\!\mathbf{R}_K\!,\!\! \, \,\! ... \!\, 
% ,\!\mathbf{R}_1\!,\!\! \, \mathbf{R}_2\!,\!\! \, ...\! \, ,\!\mathbf{R}_K\!,\!\! \, \,\! ... \!\,
% ,\!\mathbf{R}_1\!,\!\! \, \mathbf{R}_2\!,\!\! \, ...\! \, ,\!\mathbf{R}_K\!,\!\! \, \,\! ... \!\, 
% \end{bmatrix},
\end{split}
\end{equation}
Also define $\widehat{\mathbf{R}}_{N_R K T \times N_R K T}$ as a block diagonal matrix where blocks $\mathbf{R}_1\!, \mathbf{R}_2, ..., \mathbf{R}_K$ repeat on its diagonal for $T$ times, i.e., 
\begin{equation}
\mathbf {\widehat{R}}=\text{blockdiag}\{\mathbf{R}_1,\mathbf{R}_2,...,\mathbf{R}_K,...,\mathbf{R}_1,\mathbf{R}_2,...,\mathbf{R}_K\}.
\end{equation}
We use the LLR of  (\ref{Eq:LRT_movingAve1}) to get the detector for Case (3,3) as
\begin{equation}
\Lambda_{(3,3)} \!(\mathbf{V}) \!\!=2\operatorname{Re}\{\widehat{\bm \alpha}^H\widehat{\mathbf{R}}^{-1}\widehat{\mathbf v}\}-\widehat{\bm \alpha}^H\widehat{\mathbf{R}}^{-1}\widehat{\bm \alpha}. 
\end{equation}
Since the signal is unknown we use the MLE of $\widehat{\bm \alpha}$ which is $\widehat{\bm \alpha}^*=\widehat{\mathbf{R}}\widehat{\mathbf{R}}^{-1}\widehat{\mathbf{v}}=\widehat{\mathbf{v}}$. This is equivalent to replacing the MLE of $\bm \alpha_{k,t}$ into \eqref{Eq:logLikelihoodForGLRTwithMLEbased1}. Hence, the detector is written as 
\begin{equation}\label{Eq:MaxLikelihoodBaselineDet1}
\Lambda_{(3,3)}^* \!(\mathbf{V}) \!\!=\widehat{\mathbf{v}}^H \widehat{\mathbf{R}}^{-1} \widehat{\mathbf{v}}=\sum_k \sum_t \mathbf{v}_{k,t}^H \mathbf{R}_k^{-1} \mathbf{v}_{k,t}.
\end{equation}
We now present our analysis for $\Lambda_{(3,3)}^* \!(\mathbf{V})$. We write $\mathbf{R}_k^{-1}$ in (\ref{Eq:MaxLikelihoodBaselineDet1}) as $\mathbf{R}_k^{-\frac{1}{2}}\mathbf{R}_k^{-\frac{1}{2}}$. Then, the detector under $\mathcal{H}_1$, as in previous analysis, is given as
\begin{equation}\label{Eq:BaselineDetPrrofFinalSummation}
\Lambda^*_{(3,3)}|\mathcal{H}_1\!\!=\sum_t \sum_k (\mathbf{x}_{k,t} \!+\! \tilde{\mathbf{v}}_{k,t} )^H(\mathbf{x}_{k,t} \!+\! \tilde{\mathbf{v}}_{k,t}),
\end{equation}
where $\mathbf{x}_{k,t}=\mathbf{R}_k^{-\frac{1}{2}}\bm \alpha_{k,t}$ is a vector whose $j$-th element is denoted by $x_{j,k,t}$. For fixed indices $k$ and $t$, each term in the summation in (\ref{Eq:BaselineDetPrrofFinalSummation}) is a non-central chi-square $\chi^2_{2 N_R}(\sum_{j=1}^{N_R} \operatorname{Re}\{x_{j,k,t}\}^2+\operatorname{Im}\{x_{j,k,t}\}^2)$. Thus, (\ref{Eq:BaselineDetPrrofFinalSummation}) is an addition of non-central chi-squares, resulting in \begin{equation}\label{Eq:BaselineDet1TheoremH_1}
\Lambda_{(3,3)}^*|\mathcal{H}_1\!\!=\!\! \chi^2_{2 N_R K T}\!\bigg(\!\!\sum_t\! \sum_k \!\sum_{j=1}^{N_R}\! \operatorname{Re}\{x_{j,k,t}\}^2\!+\!\operatorname{Im}\{x_{j,k,t}\}^2\!\!\bigg).
\end{equation}

Similarly, $\Lambda_{(3,3)}^*|\mathcal{H}_0$ is obtained by replacing $\mathbf v_{k,t}$ in (\ref{Eq:MaxLikelihoodBaselineDet1}) with $\mathbf{R}_k^\frac{1}{2}\tilde{\mathbf{v}}_{k,t}$ and following a similar process as above. As a result, $\Lambda_{(3,3)}^* \!(\mathbf{V})$ under $\mathcal{H}_0$ is a chi-square as
\begin{equation}\label{Eq:BaselineDet1TheoremH0}
\Lambda_{(3,3)}^*|\mathcal{H}_0\!=\! \chi^2_{2 N_R K T}.
\end{equation}

\section{Upper Bound on Detection Performance}

We have assumed an unknown signal throughout our derivations and analysis thus far. In this section, we derive a detector that gives an upper bound on the target detection performance. To obtain such a detector, we assume that the signal $\bm \alpha_{k,t}$ is known. Hence, using the LLR of (\ref{Eq:LRT_movingAve1}), the upper bound detector is given as $\Lambda_{\text{Up.B.}} \!(\mathbf{V})=\operatorname{Re}\{\sum_k \sum_t \bm{\alpha}^H_{k,t}\mathbf{R}^{\!-1}_k\mathbf{v}_{k,t}\}$, where the subscript Up.B. indicates the upper bound detector. 
To calculate $\Lambda_{\text{Up.B.}} \!(\mathbf{V})$ under $\mathcal{H}_1$, similar to the previous proofs, we replace $\mathbf v_{k,t}$ in $\Lambda_{\text{Up.B.}} \!(\mathbf{V})$ with $\bm{\alpha}_{k,t}+\mathbf{R}_k^\frac{1}{2}\tilde{\mathbf{v}}_{k,t}$ where $\tilde{\mathbf{v}}_{k,t} \sim \mathcal{CN}(\mathbf 0,\, \mathbf I)$. This results in
\begin{equation}\label{Eq:upperboundDetectorUnderH1}
\Lambda_{\text{Up.B.}}| \mathcal{H}_1 \!=m_u+\!\operatorname{Re}\bigg\{\!\!\!\sum_k\! \sum_t \bm{\alpha}^H_{k,t}\mathbf{R}^{\!\!-\frac{1}{2}}_k\bm \tilde{\mathbf{v}}_{k,t}\!\bigg\}\!,
\end{equation}
where $m_u=\!\! \operatorname{Re}\!\bigg\{\!\!\!\sum_k \!\sum_t\! \bm{\alpha}^H_{k,t}\mathbf{R}^{\!-1}_k\!\bm \alpha_{k,t}\!\!\bigg\}\!$ is a known scalar. The second term in (\ref{Eq:upperboundDetectorUnderH1}) is a zero-mean real Gaussian variable with variance $v_u=\frac{1}{2}\sum_k \!\sum_t\! \bm{\alpha}^H_{k,t}\mathbf{R}^{\!-1}_k\!\bm \alpha_{k,t}$. Hence, $\Lambda_{\text{Up.B}}| \mathcal{H}_1$ is a real Gaussian variable with mean $m_u$ and variance $v_u$. The upper bound detector under $\mathcal{H}_0$ is similarly calculated by replacing $\mathbf v_{k,t}$ in $\Lambda_{\text{Up.B}} \!(\mathbf{V})$ with $\mathbf{R}_k^\frac{1}{2}\tilde{\mathbf{v}}_{k,t}$. As a result, $\Lambda_{\text{Up.B}}| \mathcal{H}_0$ is a zero-mean real Gaussian variable with variance $v_u$.

Denote by $P_{D,u}$ and $P_{FA,u}$ the probabilities of detection and false alarm for the upper bound detector. Referring to Section \ref{Sec:preliminaries}, we calculate $P_{D,u}$ in terms of $P_{FA,u}$ as $P_{D,u}=Q(Q^{-1}(P_{FA,u}) - m_u/\sqrt{v_u} )$, where $Q(.)$ indicates the Q-function.

\section{Numerical Results}\label{Sec:NumericalResults}

In this section, we validate our analysis through simulations and present a performance comparison of the detectors.

We consider a colinear configuration with either tight coupling (TC) or weak coupling (WC) for the receiver antenna array for which $N_R=32$. TC and WC respectively refer to large and small mutual coupling among the antennas, which determines the ratio $\frac{\delta}{a_R}$ where $\delta$ is set to 0.5 cm \cite{SWmassiveMIMO,SachithaPaper}. For arrays with TC, $\frac{\delta}{a_R}$ is set to the optimal value in terms of the achievable rate, i.e.,$(\frac{\delta}{a_R})^*$ which is 1.932, and for arrays with WC $\frac{\delta}{a_R}$ is set to twice the optimal value \cite{SWmassiveMIMO}. Note that superscript $*$ here indicates the optimal value. The array size is fixed and only $a_R$ changes for the two coupling cases. Our plots show detection performance for echo signals received by the sensing receiver from the end-fire direction. Note that end-fire gives the largest bandwidth gain for TCAs with a colinear configuration \cite{Bandara_2025_multiuser}. Other directions give reduced performance, with the lowest performance at broadside. All parameters related to the RLC circuit network including the resistance of the source voltage generator, the noise figure, the internal resistance of the low-noise amplifier and its gain are set as in \cite{SWmassiveMIMO}. Also, the mutual impedance among the antennas and the self-impedance of the equivalent RLC circuit for the Chu's CMS antenna are computed using \cite[Eq. (5)]{SWmassiveMIMO} and \cite[Eq. (7)]{SWmassiveMIMO} respectively. For one time sample, the  average received signal-to-noise ratio (SNR) across frequency for $K$ frequency samples is denoted by $\overline{\gamma}_K$. 

We adopt a simple signal model that captures both frequency and time variations via two easily interpretable parameters. This simple model is chosen as the purpose of this paper is to compare the performance of different detectors in different environments rather than to investigate a particular sensing scenario. Also, we assume that the signal, which models the echo signal from the target, has the same bandwidth as the transmit signal. The bandwidth of the signal is restricted to a certain range we specify for our results. The unknown signal is modeled as 
\begin{equation}\label{Eq:unknownSignalForSimulation}
\bm \alpha_{k,t} = \mathbf{h}_{k,t} g_{k,t}e^{j(k-1)\upsilon}e^{j(t-1)\tau},  
\end{equation}
where $\mathbf{h}_{k,t}$ denotes the physically-consistent channel model between the target and the receiver and $g_{k,t}$ that is the echo signal from the target is modeled as a source voltage. We consider a physically-consistent line-of-sight (LoS) channel for $\mathbf{h}_k$ given in \cite{SWmassiveMIMO} and a fixed source voltage $g_{k,t}=g$ obtained by equal power allocation among the frequency samples for one time sample. Also, in (\ref{Eq:unknownSignalForSimulation}) $\upsilon$ and $\tau$ control the rate of phase variation across frequency and time respectively. Our model has no amplitude variation except amplitude variation due to the coupling effect caused by $\mathbf{h}_{k,t}$, as variation over frequency and time is likely to be more pronounced than amplitude variation especially in a LoS channel.

Increasing $\upsilon$ induces larger frequency variations, however $\upsilon=0$ does not necessarily indicate that the signal is fixed in the frequency domain as $\mathbf{h}_{k,t}$ in (\ref{Eq:unknownSignalForSimulation}) is a frequency-dependent channel model. 
% One could get a fixed signal in the frequency domain, for a fixed bandwidth, by setting $K$ to large enough values so that frequency samples get close, effectively making $\mathbf{h}_{k,t}$ fixed across frequency.
% In our numerical results to model the signal in the frequency domain ranging from fixed to rapidly varying, large enough frequency samples are considered, indicating that $\upsilon=0$ is corresponding to a fixed signal in the frequency domain.
It is also assumed that the samples are collected in a short span of time, which essentially means that $\mathbf{h}_{k,t}$ is constant in time. Hence, $\tau=0$ does correspond to a constant signal in the time domain and $\tau$ larger than 0 causes time variations. 
   
% $\mathbb E_{K}\left[\overline{\gamma}_K\right]$.

\begin{figure}[t]
  \centering
  \begin{subfigure}[b]{0.8\linewidth}
    \includegraphics[width=\linewidth]{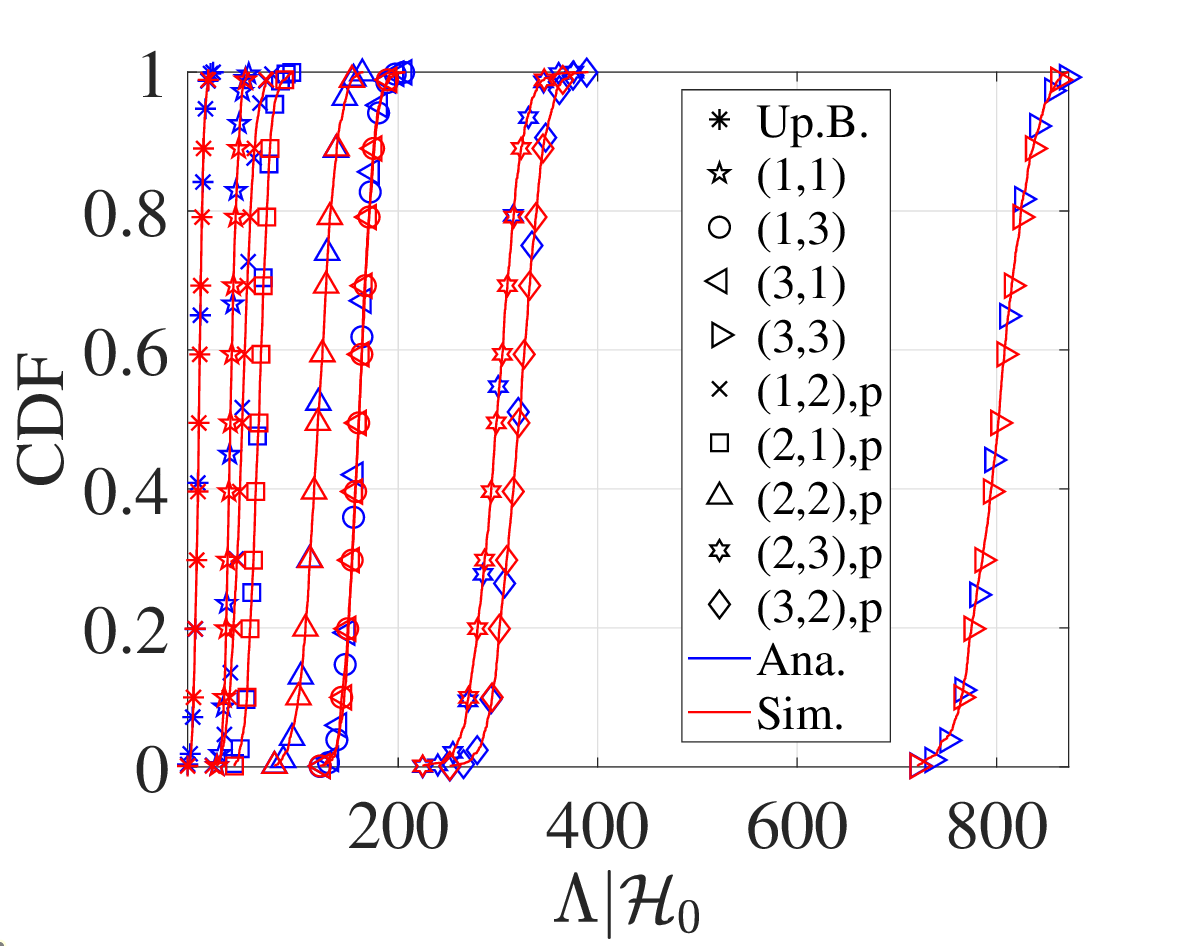}
    \caption{}
    \label{fig:a}
  \end{subfigure}
  \hfill
  \begin{subfigure}[b]{0.8\linewidth}
    \includegraphics[width=\linewidth]{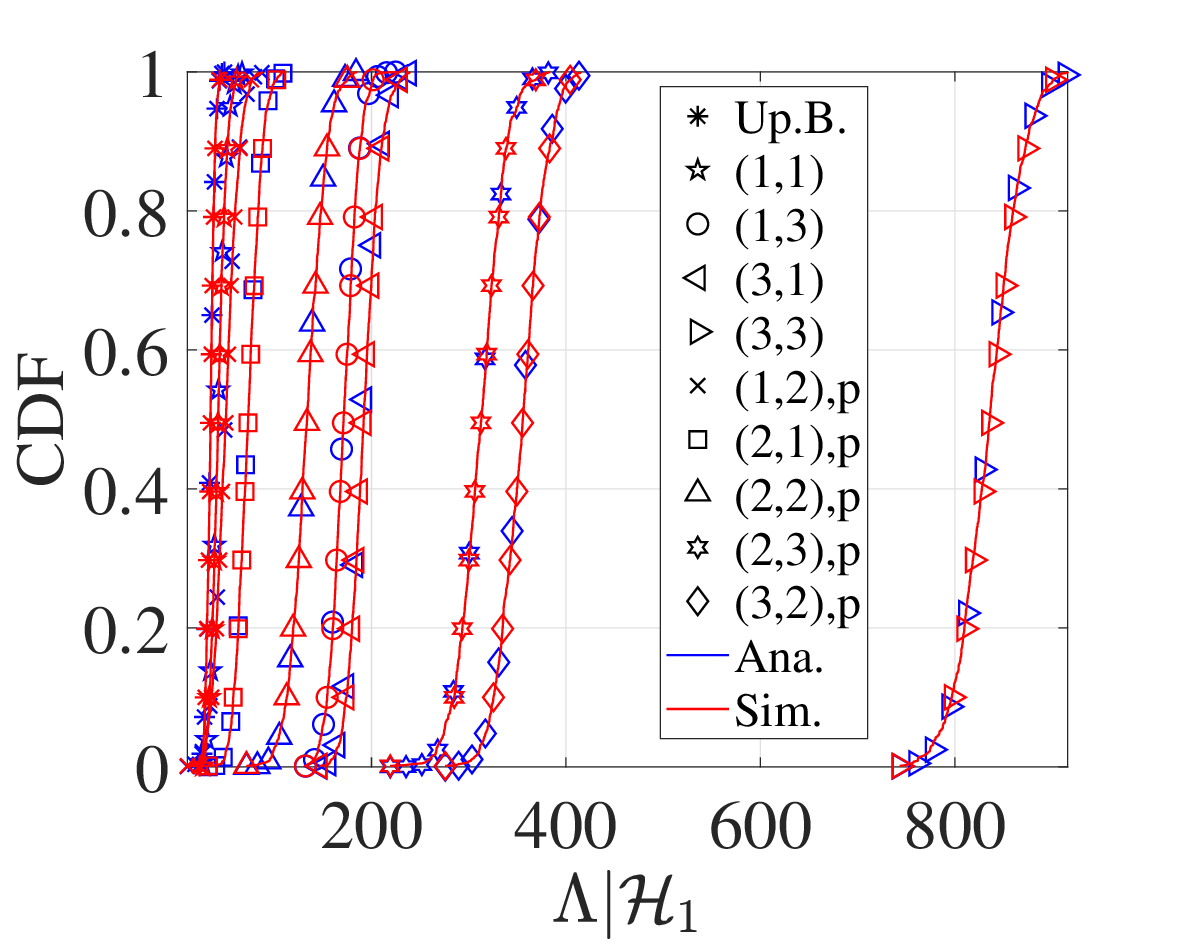}
    \caption{}
    \label{fig:b}
  \end{subfigure}
  \caption{Distributions of the detectors for (a) the null hypothesis and (b) the alternative hypothesis for TC, $K=T=5$, bandwidth [20-30] GHz, $\upsilon$ = 0.2, $\tau$ = 0.2 and $\overline{\gamma}_K$ = 1.4 dB. All window lengths are 3.}
  \label{Fig:Distrib_CompareAnaSim}
\end{figure}

\begin{figure}[t]
    \centering
    \begin{subfigure}[b]{0.75\linewidth}
        \centering
        \includegraphics[width=\linewidth]{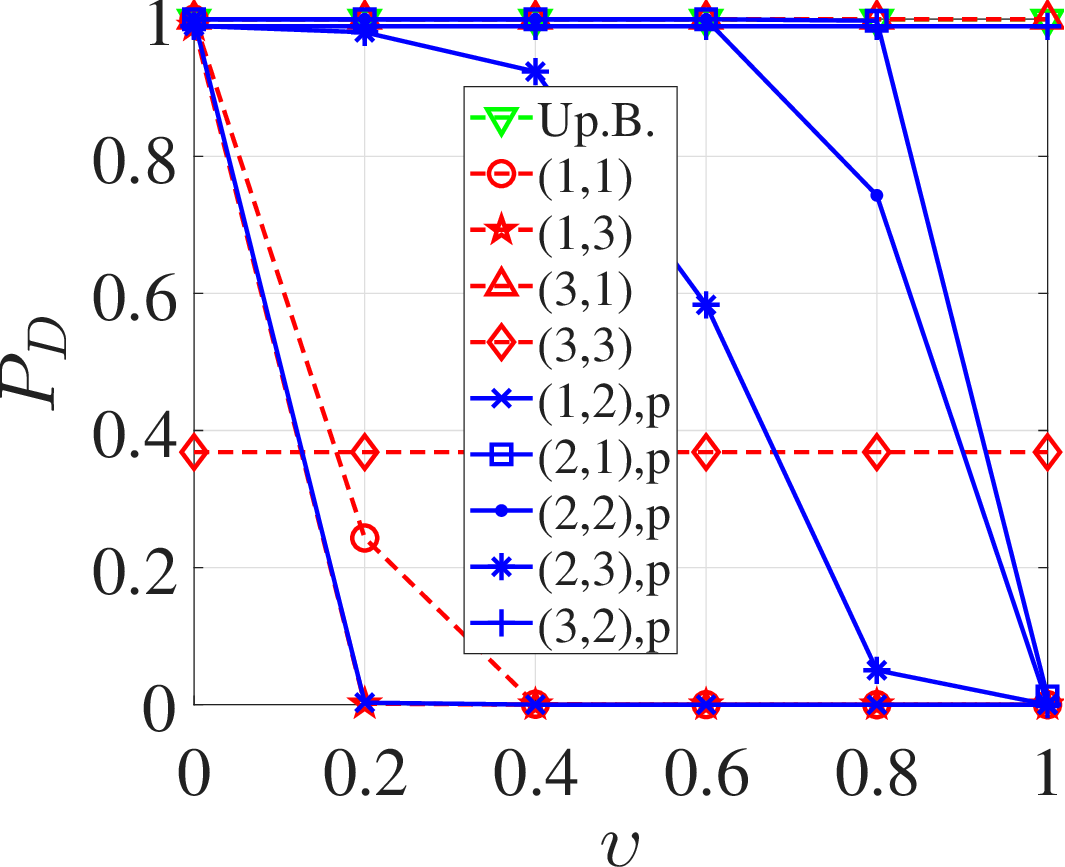}
        \caption{$\tau=0$}
        \label{Subfig:TC_Theta_tZero}
    \end{subfigure}
    % \hspace{0.001\linewidth}
    \begin{subfigure}[b]{0.75\linewidth}
        \centering
        \includegraphics[width=\linewidth]{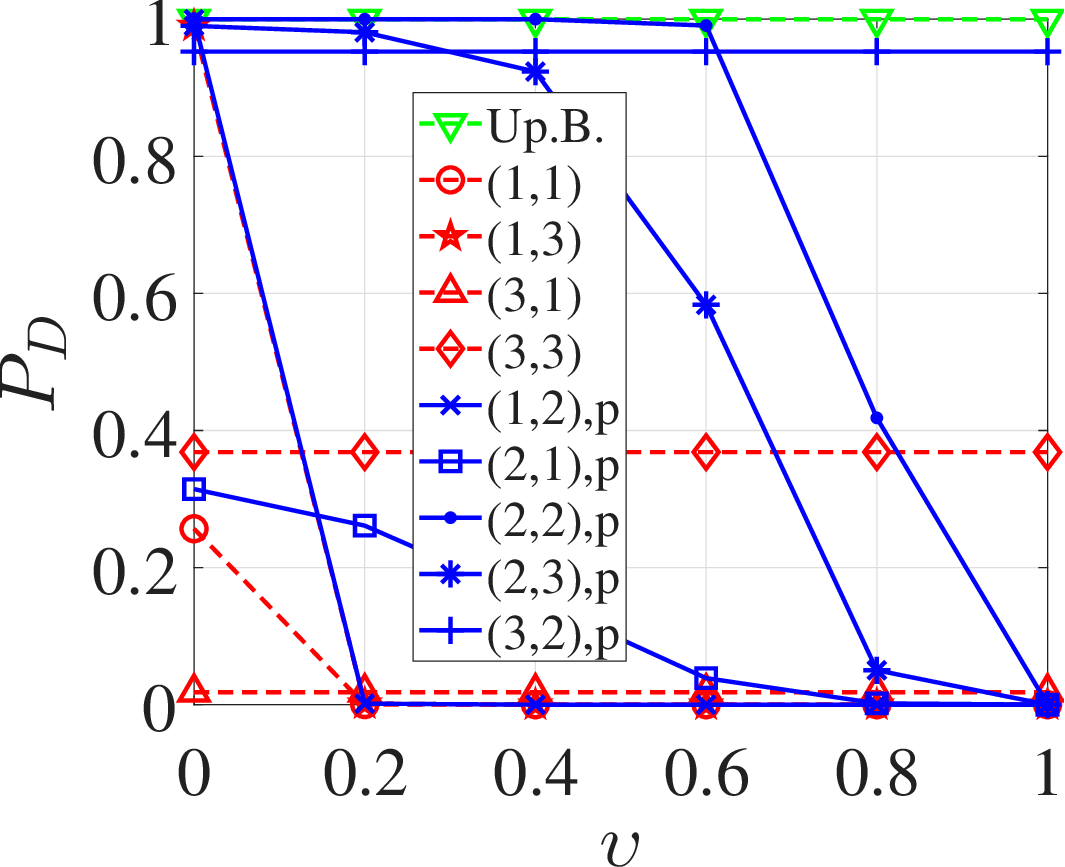}
        \caption{$\tau=0.5$}
        \label{Subfig:WWC_Theta_tZero}
    \end{subfigure}
        \begin{subfigure}[b]{0.75\linewidth}
        \centering
        \includegraphics[width=\linewidth]{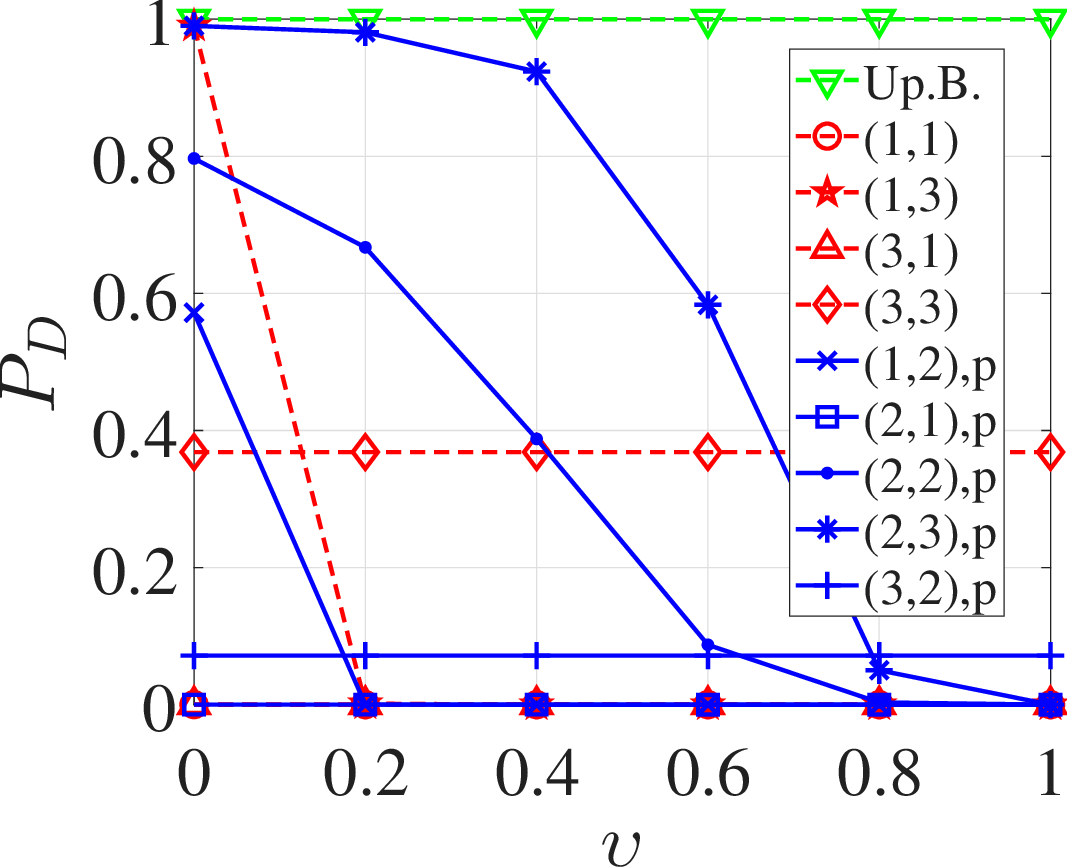}
        \caption{$\tau=1$}
        \label{Subfig:WWC_Theta_tZero}
    \end{subfigure}
    \caption{Probability of detection VS. rate of variation in frequency for an array with tight coupling (TC) with (a) $\tau=0$ (b) $\tau=0.5$ and (c) $\tau=1$, for bandwidth [28-30] GHz. Average of $\overline{\gamma}_K$ is 3.7 dB, $P_{FA}=10^{-6}$, $K=15$ and $T=8$. All window lengths are 5.}
    \label{Fig:CompareDynamics}
\end{figure}

\begin{figure}[t]
    \centering
    \includegraphics[width=0.75\linewidth]{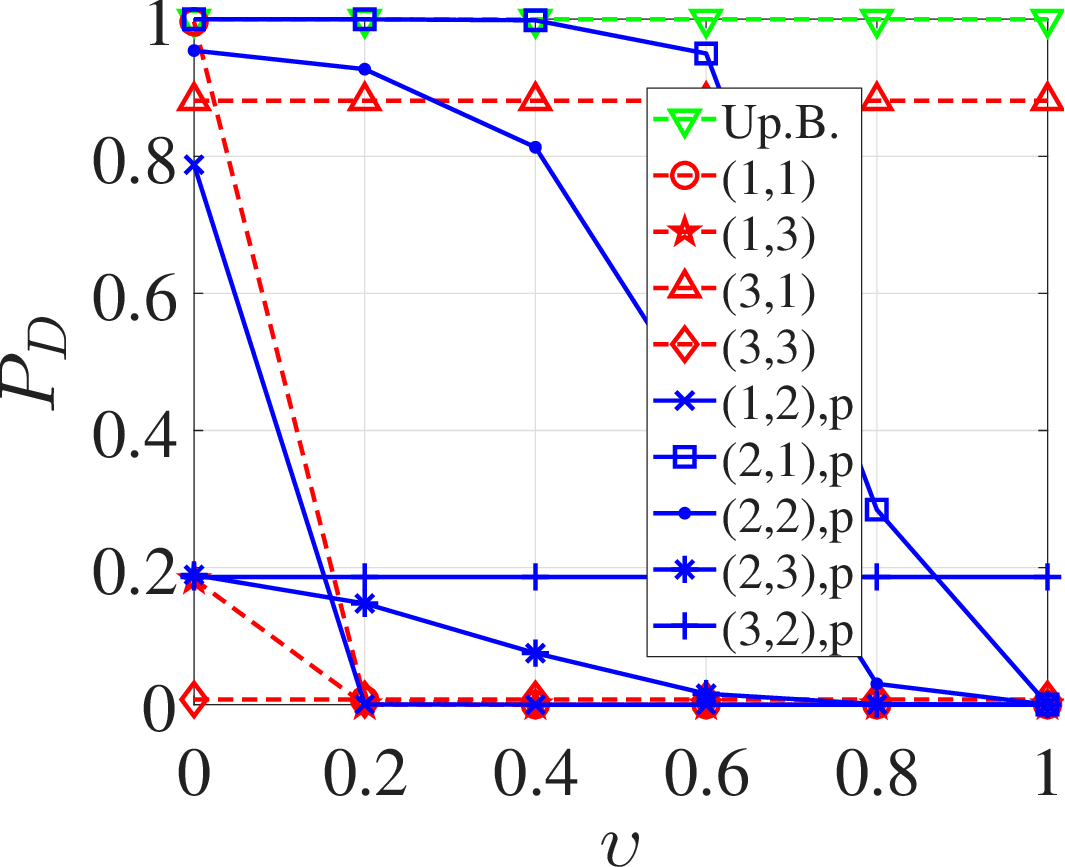}
    \caption{Detection performance for an array with weak coupling (WC) for $\tau=0$ and bandwidth [28-30] GHz. Average of $\overline{\gamma}_K$ is 0.8 dB, $P_{FA}=10^{-6}$, $K=15$ and $T=8$. All window lengths are 5.}\label{Fig:WeakCoupThetaT0}
\end{figure}

\begin{figure*}[t]
    \centering
    \subfloat[$\tau=1$\label{fig:a}]{
        \includegraphics[width=0.49\textwidth]{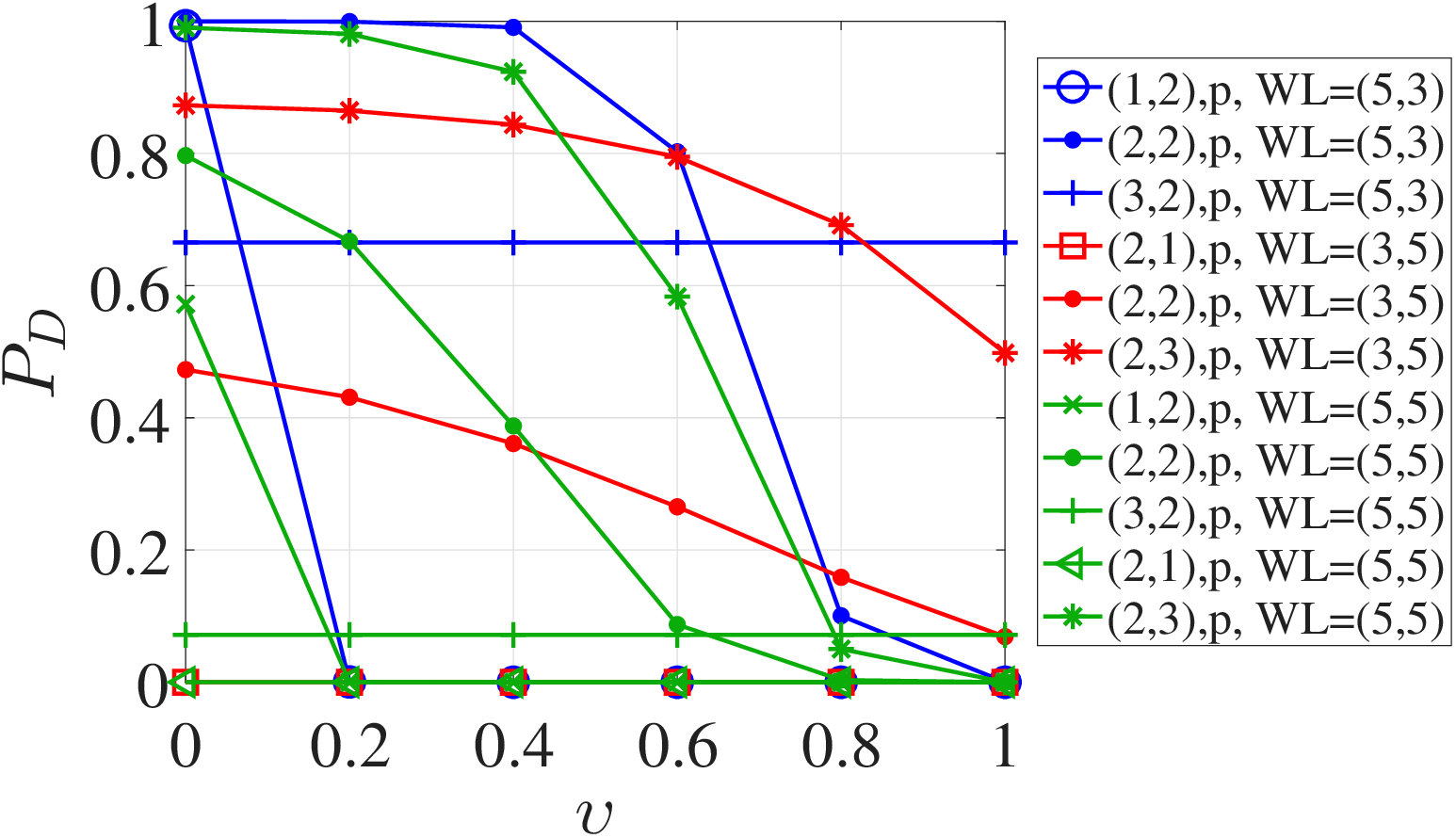}
    }
    \subfloat[$\tau=0.2$\label{fig:b}]{
        \includegraphics[width=0.49\textwidth]{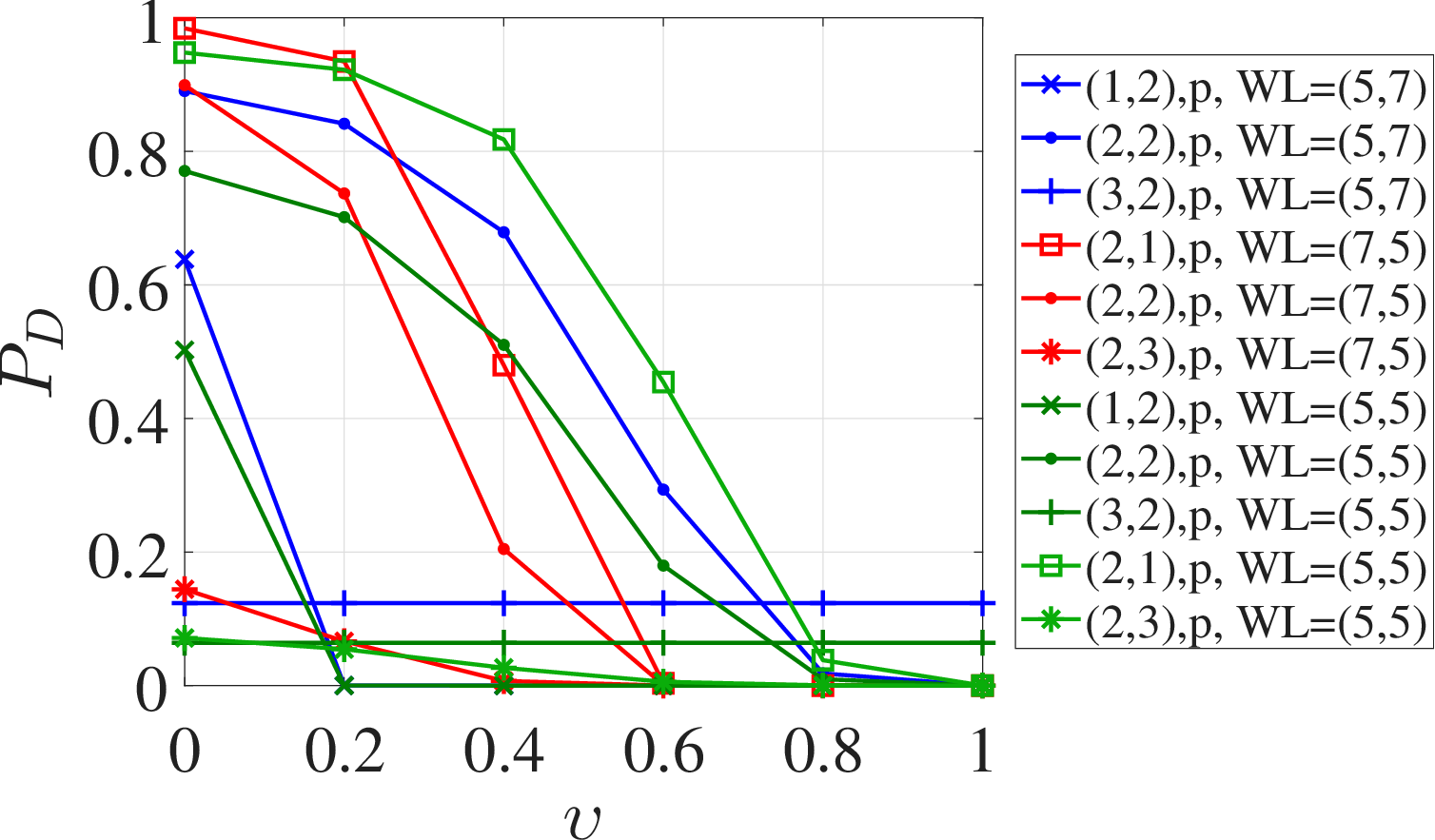}
    }
    \caption{Impact of the window lengths for an array with tight coupling (TC). (a) For rapid variation in the time domain with $\tau=1$ and average of $\overline{\gamma}_K$ is 3.7 dB. (b) For slow variation in the time domain with $\tau=0.2$ and average of $\overline{\gamma}_K$ is 0.02 dB. The bandwidth is [28-30] GHz, $P_{FA}=10^{-6}$, $K=15$ and $T=8$.}
    \label{Fig:WindowingLengthImpact}
\end{figure*}

Figs. \ref{Fig:Distrib_CompareAnaSim} (a) and \ref{Fig:Distrib_CompareAnaSim} (b) show a comparison between our analysis and simulations for the CDF of the detectors under null and alternative hypotheses. In the figure, blue markers are used to show curves corresponding to analysis. Also, red solid lines with the markers correspond to simulations. For the upper bound detector we use the abbreviation Up.B. in figure legends. For simplicity, $K$ and $T$ are set to 5. The figure shows that our analytical distributions perfectly match the simulated distributions.

Next, we show the performance of the detectors. In practice, signals are not precisely divided into the 9 possible cases in frequency and time. Thus, the assumptions that lead to the 9 detectors cannot be satisfied exactly and the detectors can be employed in scenarios for which they are not designed. Hence, it is necessary to compare the detectors in terms of robustness to spectral and temporal variation of the signal. Figs. \ref{Fig:CompareDynamics} (a)-(c) show the probability of detection $P_D$ for a fixed probability of false alarm $P_{FA}=10^{-6}$ and a large bandwidth of 2 GHz. Values of $P_D$ are computed as explained in Section \ref{Sec:preliminaries}. The letter `p' in the legend indicates our proposed detectors. In Figs. \ref{Fig:CompareDynamics} (a)-(c), the x-axis shows the rate of variation in frequency, i.e. $\upsilon$. Figs. \ref{Fig:CompareDynamics} (a)-(c) each have a different $\tau$ value to show the impact of the time variation rate, with $\tau$ set to 0, 0.5 and 1 respectively. Also, we set the total power such that the upper bound detector approaches $P_D=1$, ensuring the other detectors are not restricted by limited power. Note that the window lengths in frequency and time for our proposed detectors are fixed across Figs. \ref{Fig:CompareDynamics} (a)-(c). This corresponds to the worst case performance of our detectors as the performance of our proposed detectors can be improved by optimizing the window lengths based on the specific scenario. We show the impact of the window length later in this section. As seen, even in the worst case, our detectors are robust across different scenarios and show the best overall performance. This means that, as illustrated, other detectors work only in specific scenarios but in all scenarios a suitable proposed detector works equally well or better. The only exception is the detector for Case (3,3) for $\upsilon=1$ in Fig. \ref{Fig:CompareDynamics} (c) that shows higher $P_D$, but we emphasize that the figure shows the worst case performance of our detectors. By optimizing the window length, our proposed detectors result in performance superior to that of the detector for Case (3,3).

In Figs. \ref{Fig:CompareDynamics} (a)-(c), detectors that are designed for rapid variations in the frequency domain, i.e., detectors for Cases (3,1), (3,2) and (3,3), have constant $P_D$. This is because they work similarly to an energy detector in the frequency domain, with phase changes or changing $\upsilon$ having no effect on their performance. Also, phase changes in the time domain 
has no impact on the detection performance of the detectors designed for rapid variations in the time domain, that is the detectors for Cases (1,3), (2,3) and (3,3). These detectors show the same performance for all values of $\tau$. This is evident by comparing Figs. \ref{Fig:CompareDynamics} (a)-(c). In addition, for the bandwidth considered, the channel $\mathbf{h}_{k,t}$ varies slightly in frequency. As a result, the detectors for Cases (1,1), (1,2) and (1,3) perform better for very small $\upsilon$ compared to larger $\upsilon$.

Fig. \ref{Fig:WeakCoupThetaT0} shows detection performance for an array with WC for $\tau=0$. The merit of TC in detection is clear by comparing Fig. \ref{Fig:WeakCoupThetaT0} with Fig. \ref{Fig:CompareDynamics} (a) plotted for an array with TC. Even though for both the arrays the same total power of the echoes is used, the average SNR over the considered wide bandwidth for the array with TC is larger than that of for the array with WC. The larger SNR value across the bandwidth for arrays with TC is due to the coupling effect as shown in \cite{SWmassiveMIMO, SachithaPaper}. This contributes to better detection performance of the array with TC. In fact, the detection performance for every detector for the array with WC degrades, except for the upper bound detector.
For brevity we omit showing results for $\tau$ larger than 0, since the same observation applies.

Fig. \ref{Fig:WindowingLengthImpact} (a) illustrates the impact of window length on the detection performance of our proposed approach for rapid variations in the time domain with $\tau=1$. To have a concise legend for the figure, we define $\text{WL}_\text{k}$ and $\text{WL}_\text{t}$ as the window lengths in the frequency and time domains that are related to the window parameters in Section \ref{Sec:ProposedForAtleastOneSlow2} as $L_2=L_3=M=(\text{WL}_\text{k}-1)/2$ and $L_1=L_4=Q=(\text{WL}_\text{t}-1)/2$. We also define the pair $\text{WL}=(\text{WL}_\text{k},\text{WL}_\text{t})$ which is shown in the legend.

Fig. \ref{Fig:WindowingLengthImpact} (a)  shows that when there is large variation in either frequency or time domains smaller window lengths yield a better performance. The figure is plotted for $\tau=1$ indicating rapid variation across time. As can be seen, the curves with $\text{WL}=(5,3)$ perform better than the curves with $\text{WL}=(5,5)$. Also, it can be seen that curves with $\text{WL}=(3,5)$ shows better performance than the curves with $\text{WL}=(5,5)$ when $\upsilon$ is increased. 
%For faster variations, the weighted average that estimates the signal gets inaccurate by increasing the window lengths, thus smaller window lengths contribute to better performance.

Fig. \ref{Fig:WindowingLengthImpact} (b) shows how window length affects the detection performance of our proposed approach for slow variations in the time domain with $\tau=0.2$. For this figure, a smaller total power is used to show the distinction between the detectors when $\upsilon$ is small. The figure shows when there is slow variation in either frequency or time domains, larger window lengths contribute to a better performance. With the slow variation in time corresponding to $\tau=0.2$, it can be seen that the curves with $\text{WL}=(5,7)$ show superior performance compared to the curves with $\text{WL}=(5,5)$. In addition, it is evident from the figure that for small $\upsilon$, the curves for $\text{WL}=(7,5)$ show better performance compared to the curves for $\text{WL}=(5,5)$. When the signal is slowly varying, increasing the window lengths results in better averaging and consequently better estimation of the signal. As seen, even a small change in $\text{WL}_\text{k}$ and $\text{WL}_\text{t}$ has a notable impact on detection performance. Of course, $\text{WL}_\text{k}$ and $\text{WL}_\text{t}$ can be optimized to make our detectors work very efficiently, however this is left for future work.

\section{Conclusion}
We conducted joint time and frequency target detection assuming unknown wideband signals and using TCAs. We considered correlated noise with a frequency-dependent covariance matrix. We provided detectors and analysis for signals with various temporal/spectral dynamics and proposed a novel detection approach for slowly-varying signals. We also derived a detector with upper bound performance. Our results demonstrate the notable benefits of TCAs over weakly-coupled antennas in target detection, caused by the higher SNR TCAs provide over a large bandwidth. We examined different detectors in practical scenarios where assumptions used to design them are not exact and highlighted the superior performance of our proposed approach. 

%In future work, the benefits of TCAs in target detection can be explored for known signal models which will be useful for ISAC waveform designs.

% We showed that our analysis matches the simulations and that our detection approach outperforms the other detectors.
% \appendices

\bibliographystyle{IEEEtran}
\bibliography{References}
\end{document}